\documentclass[reprint,superscriptaddress,showpacs,preprintnumbers,nofootinbib,amsmath,amssymb,aps,onecolumn,prd,
floatfix
]{revtex4}
\usepackage[utf8]{inputenc}
\usepackage{graphicx}
\usepackage{epsfig}
\usepackage{dcolumn}
\usepackage{bm}
\usepackage{amssymb}
\usepackage{amsmath}
\usepackage{subfigure}
\usepackage[colorlinks]{hyperref}
\usepackage[usenames,dvipsnames]{color}

\hypersetup{
      breaklinks=true,
    pdfstartview={FitH},    
    colorlinks=true,       
    linkcolor=blue,          
    citecolor=red,        
    filecolor=magenta,      
    urlcolor=blue,           
    anchorcolor=green,      
    linktocpage=true
}

\usepackage[tableposition=top]{caption}
\DeclareCaptionLabelFormat{gostfigure}{Fig. #2}
\DeclareCaptionLabelSeparator{gost}{.~~}
\begin{document}
\title{Emergence of cosmological scaling behavior in asymptotic regime}
\author{M.~A.~Skugoreva}
\email{masha-sk@mail.ru} \affiliation{Kazan Federal University, Kremlevskaya 18, Kazan, 420008, Russia}
\author{M.~Sami}
\email{samijamia@gmail.com}  \affiliation{Centre
for Theoretical Physics, Jamia Millia Islamia, New Delhi-110025,
India}\affiliation{Maulana Azad National Urdu
University, Gachibowli, Hyderabad-500032,India}\affiliation{Institute for Advanced Physics $\&$ Mathematics, Zhejiang University of Technology,\\ Hangzhou, 310032, China}
\author{N. Jaman}
\email{nurjaman@ctp-jamia.res.in} \affiliation{Centre
for Theoretical Physics, Jamia Millia Islamia, New Delhi-110025,
India}
 
\begin{abstract}
    We consider a scalar field system with a class of potentials given by the expression, $V(\phi)\propto \phi^m {\rm exp}({-\lambda \phi^n/{M^n_{Pl}}})$, $m\geqslant 0, n>1$ for which $\Gamma=V_{\phi \phi}V/V^2_{\phi}\to 1 $ as $|\phi|\to \infty$. We carry out a dynamical analysis for the underlying system choosing a suitable set of autonomous variables and find all of the fixed points. In particular, we show that the scaling solution is an attractor of the system in the asymptotic regime. We indicate the application of the solution to models of quintessential inflation.
\end{abstract}

\maketitle
\section{Introduction}
    Scalar fields have been used extensively in cosmology for the description of the inflationary era as well as the phenomenon of late-time acceleration. In their application to late-time acceleration, it is imperative that they do not interfere with the thermal history of the Universe, and the dynamics is such that the late-time physics bears no dependence on the initial conditions. The first requirement \textit{\`{a} la} the the nucleosynthesis constraint \cite{Copeland,Ferreira:1997au,Sami:2004ic,Sahni:2001qp,Tashiro:2003qp} implies that the field energy density should remain subdominant to background during radiation and matter era and should show up only at late stages to account for the late-time acceleration. In general, the first requirement asks for a steep field potential, whereas the second forces the choice of a particular type of steep behavior. In the case of the early field dominance, the field energy, during evolution, undershoots the background such that the Hubble damping freezes the scalar field $\phi$ on its potential.\footnote{In this case, the field energy density is subdominant to the background initially, the field remains frozen till its energy density becomes comparable to the background. Thereafter evolution depends upon the nature of steepness of the potential, similar to the case described in the text.} As the background energy density becomes comparable to the field energy density, field evolution commences again. Hereafter, the evolution crucially depends on the nature of steepness of the field potential. In a case where the underlying field potential is of standard exponential type, the field energy density exactly tracks the background forever \cite{Copeland}. If the potential is less steep than the exponential --- for instance, the inverse power-law potentials --- the field energy density gradually approaches the background and overtakes it. And for potentials more steep than the standard exponential, the field energy density would evolve away from the background, pushing the field into the freezing regime and after the recovery from the latter \cite{Hossain:2014zma,Geng:2015fla}. The same behavior keeps repeating till later.\footnote{We imagine a late-time feature in the potential allowing the exit to acceleration.} Such a framework belongs to the class of thawing models \cite{Linder:2008pp,thawing1}, where the evolution is sensitive to the initial conditions.

    Recently, generalized exponential potential $V(\phi)\propto {\rm exp}(-\lambda\phi^n/{{M^n_{Pl}}})$, $n>1$ was considered in Refs. \cite{Sami,Geng:2015fla} in the context of quintessential inflation \cite{Spokoiny:1993kt,Peebles:1998qn,Peebles:1999fz,Peloso:1999dm,Dimopoulos:2000md,Copeland:2000hn,Majumdar:2001mm,Sami:2004xk,Rosenfeld:2005mt,Dimopoulos:2001ix,Giovannini:2003jw,Sami:2004ic,Tsujikawa:2013fta,Hossain:2014xha,Hossain:2014zma,Dimopoulos:2017zvq,Linde, Ahmad:2017itq,Dimopoulos:2018,Jaman:2018ucm,Hossain:2018pnf,Cardenas:2006py,Sanchez1,Sanchez2,Neupane:2007mu,BasteroGil:2009eb,Guendelman:2016kwj,Haro1,Haro2}, which successfully describes inflation and mimics desirable behavior \cite{Steinhardt:1999nw,Ferreira:1997hj,Ratra:1987rm,Guo:2003eu,Copeland:2006wr,Singh} at late stages despite the fact that the slope of the potential is not constant, which is important for the derivation of the scaling solution. In this case, one has an additional equation for the slope, where a crucial role is played by a quantity $\Gamma=V_{\phi \phi}V/V^2_{\phi}$ \cite{Steinhardt:1999nw} such that the latter being equal to $1$ implies the standard case of exponential potential. There might be interesting cases in which $\Gamma$ approaches unity in certain limits, signaling the emergence of scaling behavior asymptotically; the mentioned class of potential satisfies said criteria.

    It is desirable to have scaling behavior in models of dark energy, which makes the evolution free of initial conditions~\cite{Steinhardt:1999nw}. However, since the scaling solutions are not accelerating, one needs a late-time exit from scaling regime to acceleration which can be triggered using various mechanisms discussed in the literature \cite{Hossain:2014xha,Wetterich:2013jsa}. Scaling behavior with a mechanism of late-time exit to acceleration dubbed a {\it tracker} \cite{Steinhardt:1999nw,Caldwell:1997ii,trac1,trac2,Ng,Chiba:2009gg} is at the heart of model building for dark energy and quintessential inflation.

    In this paper, we consider a scalar field system with a generalized class of potentials with the desired property of $\Gamma$ and carry out a dynamical analysis rigorously to confirm the existence and stability of scaling solutions using the autonomous system framework; the analysis was missing in the earlier cited work. We also indicate the applications of these solutions for models of quintessential inflation. 
    
\section{Dynamical analysis}
    In what follows, we shall present the evolution equations in the autonomous form suitable for the study of fixed points. In particular, our focus will be on the existence and stability of the scaling solution, which plays an important role in model building of dark energy and quintessential inflation. Besides the autonomous form, we shall also retain evolution equations in the original variables, which would be helpful in constructing certain physical quantities along with the additional check on the results to be obtained from the autonomous system. We shall use units $\hbar=c=1$.
    
\subsection{Equations of motion}
    Let us consider the model of quintessence with the following action,
\begin{equation}
\label{action}
S=\int{d^4x\sqrt{-g}\left[-\frac{{M_{Pl}}^2}{2}R+\frac{1}{2}{(\nabla\phi)}^2
-V(\phi)\right]}+S_M,
\end{equation}
 where
\begin{equation}
\label{pot}
V(\phi)=V_0{\left(\frac{\phi}{M_{Pl}}\right)}^m e^ {-\lambda\frac{\phi^n} {{M_{Pl}}^n}},~~~\lambda>0, V_0>0, n>1, m\geqslant0
\end{equation}
is the scalar field potential, ${{M_{Pl}}^2}=\frac{1}{8\pi G}$ and $S_M$ is the matter action. The case with  $m=0$ and $n=1$ has been thoroughly studied in the literature \cite{Sahni:2001qp,Copeland:2000hn,Hossain:2014xha,Lucchin:1984yf,Sami:2002fs,Ratra:1989uz,Joyce}. Several asymptotic solutions were found in Refs.~\cite{Ng, Barrow1} in the models with various parameters $n$, $m$ of the potential (\ref{pot}). A mixed-quintessence potential similar to Eq.~(\ref{pot}) appears in the Einstein-frame formulation of the nonminimal coupled scalar field models, and it is obtained in Ref.~\cite{Wetterich} that a corresponding system shows scaling behavior in spite of having a potential without constant slope.

    Hereafter, we specialize to a spatially flat Friedmann-Lema\^{i}tre-Robertson-Walker metric $ds^2=dt^2-a^2(t)\delta_{ij}{dx}^i {dx}^j$, where $a$ is the scale factor. Equations of motion are obtained by varying the action (\ref{action}) with respect to the metric and with respect to the scalar field, giving rise to the usual set of evolution equations,
\begin{equation}
\label{system1}
3H^2{M^2_{Pl}}=\frac{1}{2}{\dot\phi}^2+V(\phi)+\rho,
\end{equation}
\begin{equation}
\label{system2}
(2\dot H+3H^2)M^2_{Pl}=-\frac{1}{2}{\dot\phi}^2+V(\phi)-p,
\end{equation}
\begin{equation}
\label{system3}
\ddot\phi+3H\dot\phi+V_{\phi}=0,
\end{equation}
where $H(t)\equiv\frac{\dot a}{a}$ is the Hubble parameter, $\rho$, $p$ are the energy density and pressure of the background matter, $w=p/\rho$ is the background matter equation of state parameter ($w\in[-1; 1]$) and $V_{\phi}=\frac{dV}{d\phi}$.

    Let us note that the asymptotic scaling solution in the model with the potential $V(\phi)=V_0e^{-\lambda\frac{\phi^n} {{M_{Pl}}^n}}$ was constructed in Ref.~\cite{Sami}, though the existence and stability was not demonstrated there using dynamical system analysis. We could expect that this solution exists for a general potential $V(\phi)=V_0{\left( \frac{\phi}{M_{Pl}}\right) }^m {\rm exp}\left({-\lambda\frac{\phi^n} {{M_{Pl}}^n}}\right)$ also, as the parameter $\Gamma\equiv V_{\phi \phi}V/V^2_\phi$ has the same asymptotic behavior in this case. This is not surprising as the power law gives an insignificant contribution in the asymptotic regime. In this case, we are interested in investigating all of the fixed points, in particular, the scaling solution. 
   
   Let us note that the behavior of the scalar field in the scaling regime is characterized by 
\begin{equation}
\label{scaling1}
w_{\phi}=\frac{p_{\phi}}{\rho_{\phi}}=\frac{{\dot\phi}^2/2-V}{{\dot\phi}^2/2+V}=w ~~\Rightarrow~~ {\dot\phi}^2(1-w)=2(1+w)V,
\end{equation}
where $\rho_{\phi}$, $p_{\phi}$, $w_{\phi}$ are the energy density, the pressure and the equation of state parameter of the scalar field $\phi$, respectively. The time derivative from the last relation in Eq.~(\ref{scaling1}) is
\begin{equation}
\begin{array}{l}
\label{scaling2}
\ddot\phi\dot\phi(1-w)=(1+w)V_{\phi}\dot\phi ~~\Rightarrow~~ (-3H\dot\phi -V_{\phi})\dot\phi(1-w)=(1+w)V_{\phi}\dot\phi ~~\Rightarrow 
\\~~~~~~~~~~~~~~~~~~~~~~~~~~~~~~~~~~\Rightarrow~~ -3(1-w)H{\dot\phi}^2=2V_{\phi}\dot\phi.
\end{array}
\end{equation}
Therefore, we find for the scaling solution
\begin{equation}
\label{scaling3}
\frac{{\dot\phi}^2}{2V}=\frac{1+w}{1-w}, ~~~~\frac{V_{\phi}}{\dot\phi H}=\frac{3}{2}(w-1).
\end{equation} 

    Some of the features of the dynamics can be made clear by looking at the form of the potential. We plot the potential $V(\phi)=V_0 {\left( \frac{\phi}{M_{Pl}}\right) }^m e^{-\lambda\frac{\phi^n} {{M_{Pl}}^n}}$ for several sets of parameters, $n$, $m$, $\lambda$, $V_0$, and we observe that for $n$ even and $m$ odd, the potential has a minimum for the negative value of the field and should give rise to de Sitter solution (see Fig.~\ref{Fig2}, right plot). However, this case cannot be captured by our choice of autonomous variables; they are best suited to scaling solutions. In a case of even $m>0$ and both $n$ being even or odd, the potential has a local minimum at $\phi=0$ with $V=0$. In this case, we expect that the stable fixed point would correspond to $\Omega_{\phi}\to 0$, corresponding to $w_{\rm eff}=\frac{p+p_{\phi}}{\rho+\rho_{\phi}}\to w$. Thus, apart from the scaling solutions, it is expected that the formalism would capture the latter behavior also.
\begin{figure}[hbtp]
\includegraphics[scale=0.46]{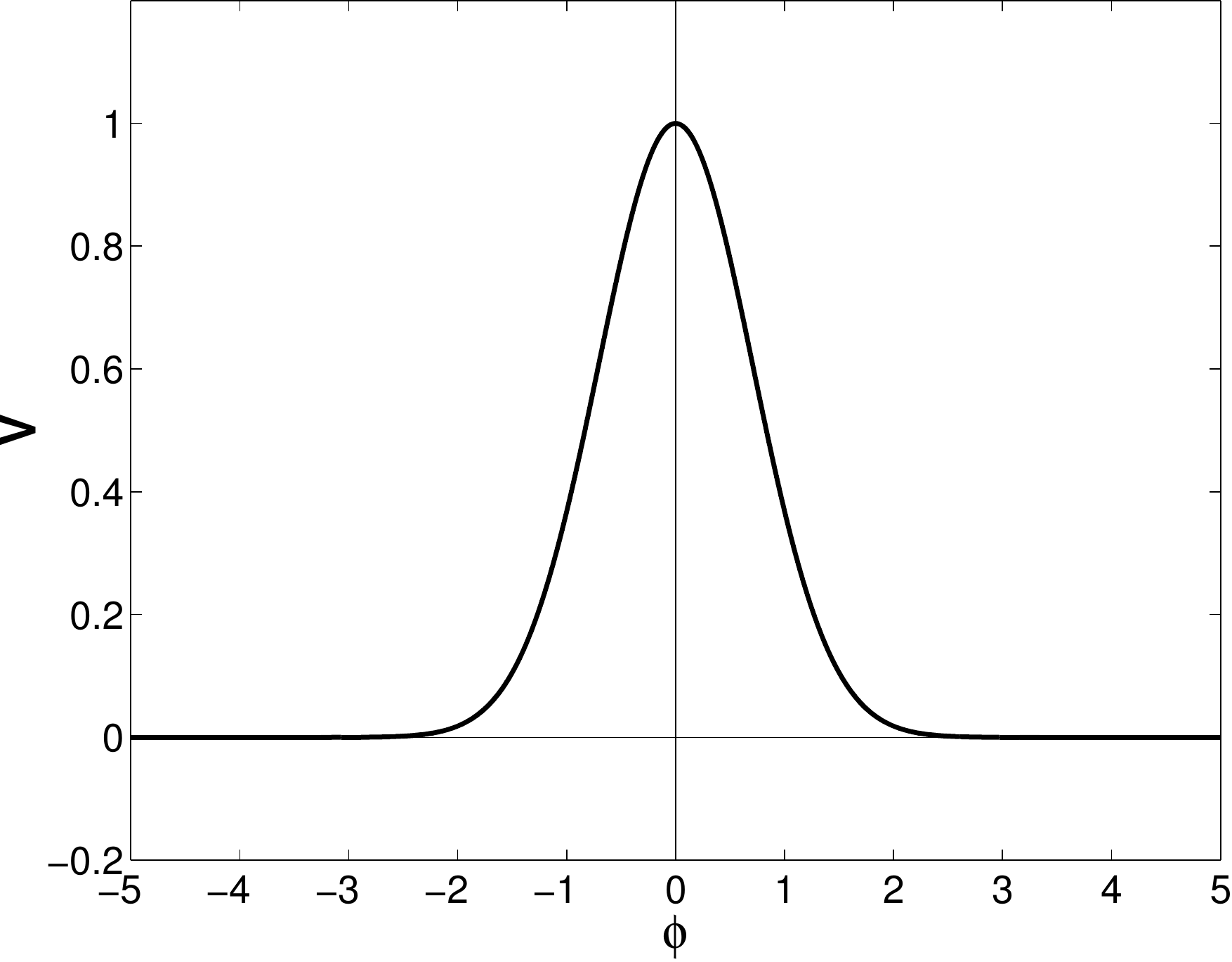}\qquad
\includegraphics[scale=0.46]{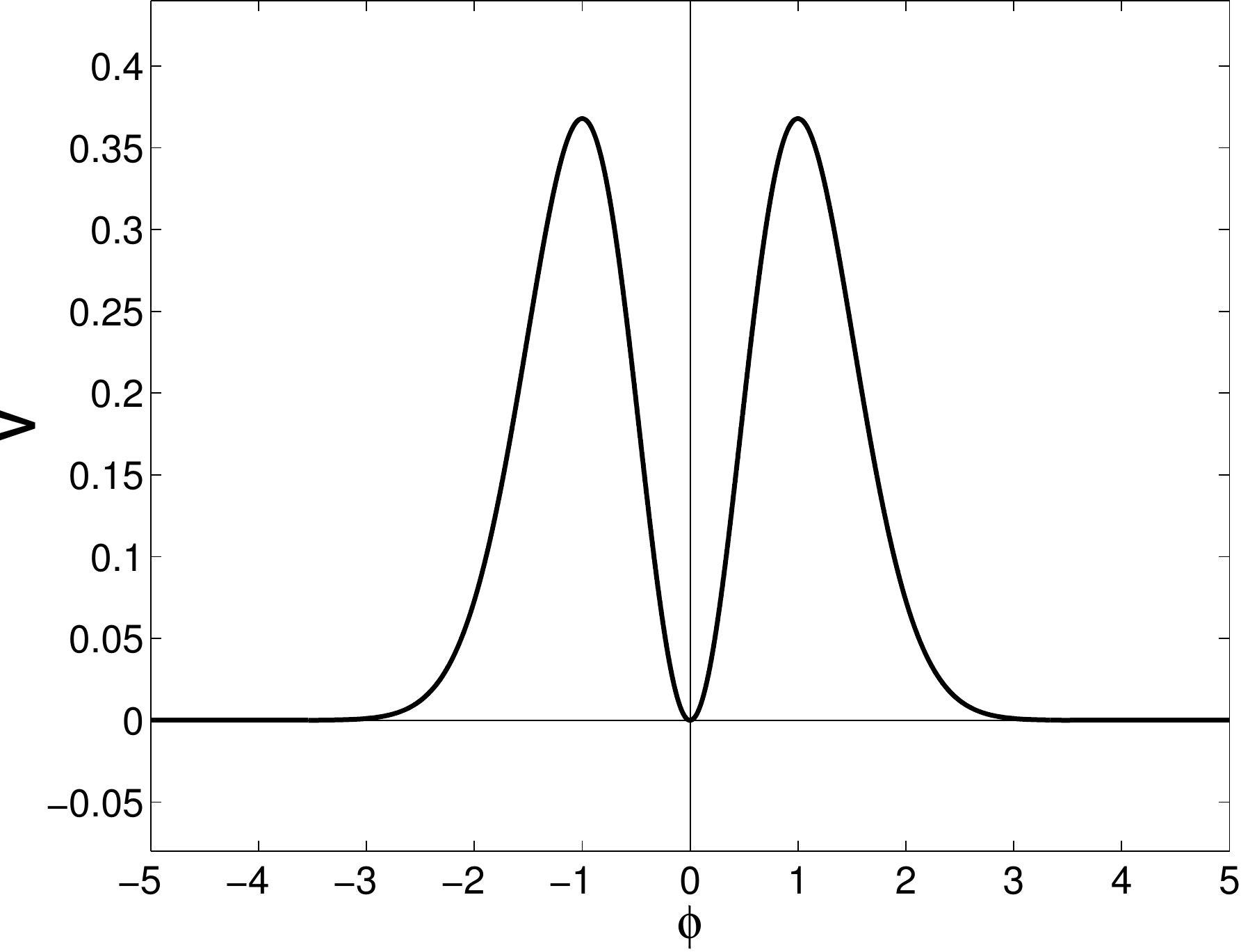}
\caption{The scalar field potential $V(\phi)=V_0{\left( \frac{\phi}{M_{Pl}}\right) }^m e^{-\lambda\frac{\phi^n} {{M_{Pl}}^n}}$ is plotted for the parameters ~~(left~panel)~$n=2$,~$m=0$~~ and ~~(right panel)~$n=2$,~$m=2$.~~ Other parameters are ~$V_0=1$, ~~$\lambda=1$, ~~${M^2_{Pl}}=1$.}
\label{Fig1}
\end{figure}
\begin{figure}[hbtp] 
\includegraphics[scale=0.32]{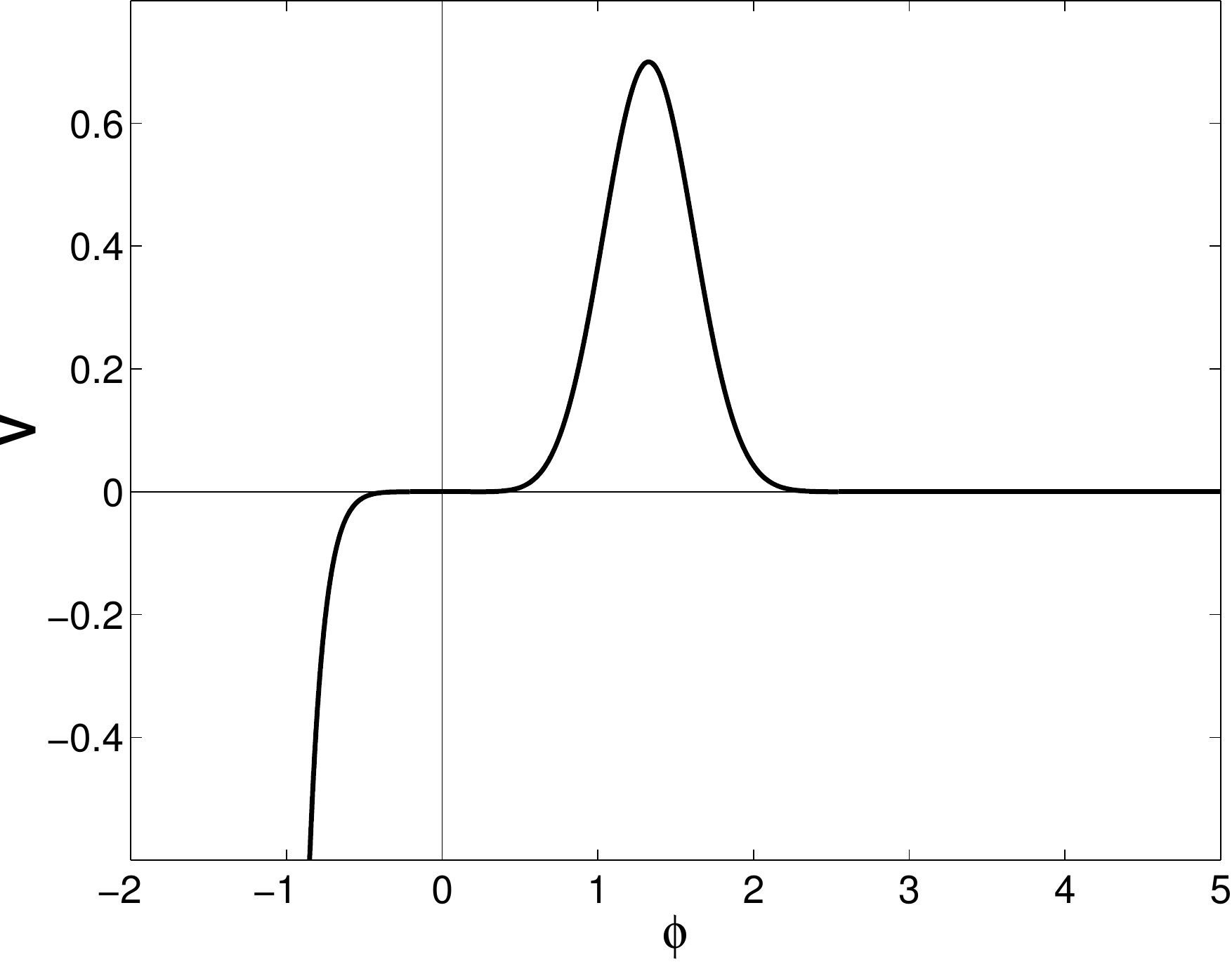}
\includegraphics[scale=0.32]{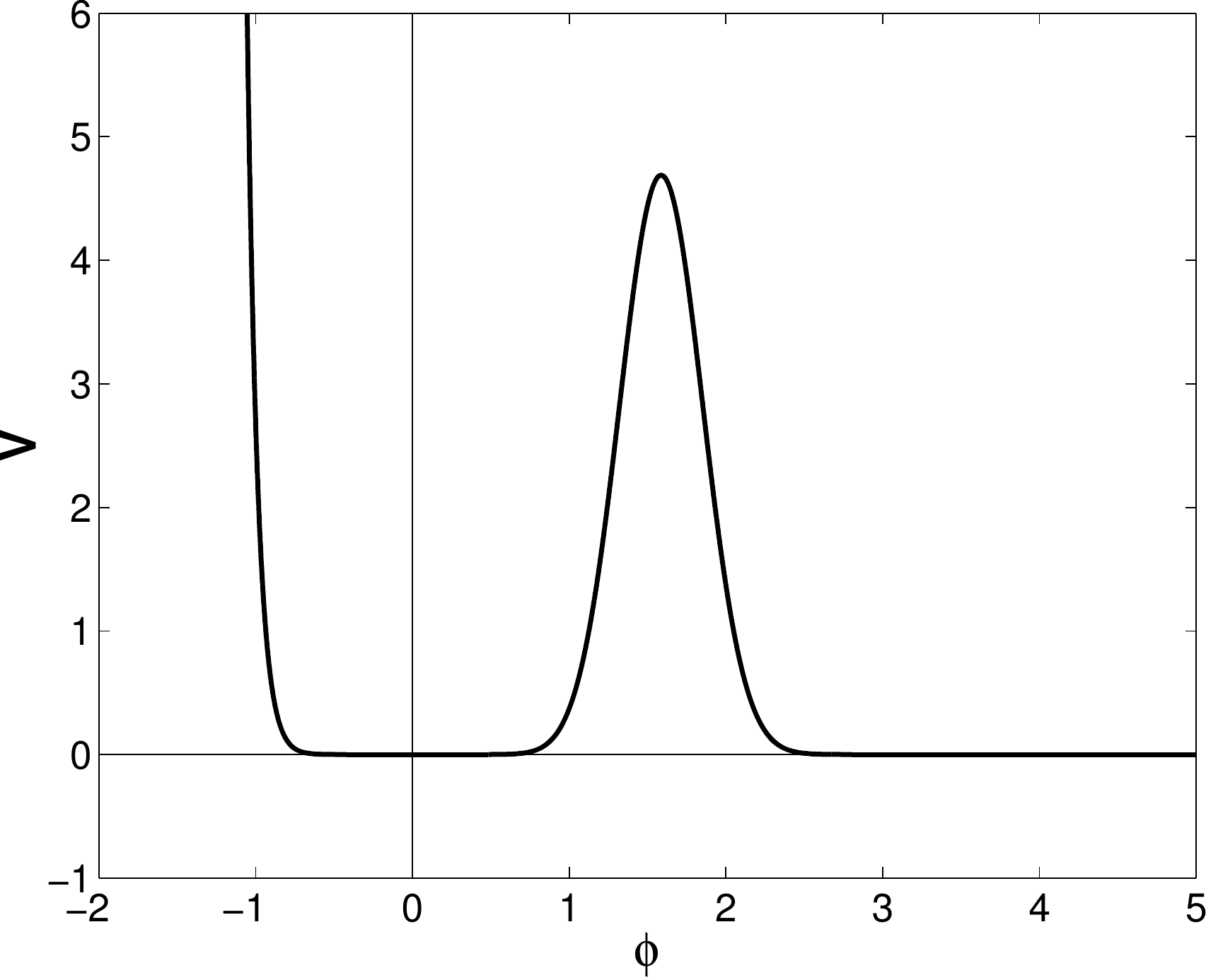}
\includegraphics[scale=0.32]{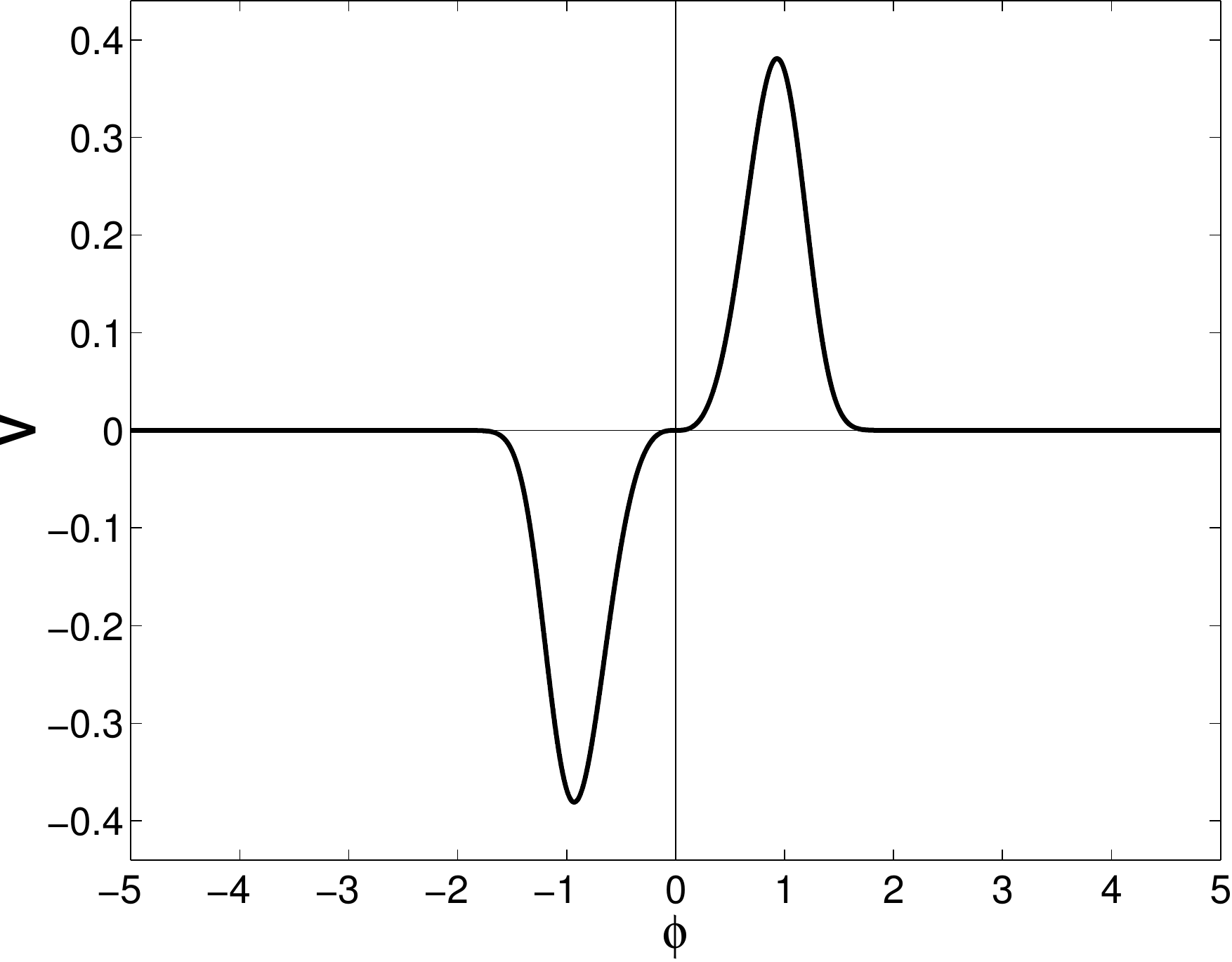}
\caption{The scalar field potential $V(\phi)=V_0{\left( \frac{\phi}{M_{Pl}}\right) }^m e^{-\lambda\frac{\phi^n} {{M_{Pl}}^n}}$ is plotted for the parameters ~~(left~panel)~$n=3$,~$m=7$,~~ (middle panel)~$n=3$,~$m=12$~~ and ~~(right panel)~$n=4$, $m=3$.~~ Other parameters are ~$V_0=1$, ~~$\lambda=1$, ~~${M_{Pl}}^2=1$.}
\label{Fig2}
\end{figure}   
      
    We introduce new expansion-normalized variables $x$, $y$, $A$:
\begin{equation}
\label{xyA}
x=\frac{{\dot\phi}^2}{2V},~~~~~~~~ y=\frac{V_{\phi}}{\dot\phi H},~~~~~~~~ A=\frac{1}{\phi/M_{Pl}+1}.
\end{equation}
Let us note that the variable $x$ is positive if $V(\phi)>0$ and the variables $x$ and $y$ in Eq.~(\ref{xyA}) are constants in the scaling regime, which happens exactly in the case of a standard exponential potential where the third relation is redundant. For a general class of potentials with a property, the parameter $\Gamma\to 1$ as $|\phi|\to \infty$, we have an asymptotic scaling regime where $A$ is zero.

    The initial variables $\phi$, $H$, $\rho$ and also their time derivatives and expressions of these variables are expressed through autonomous variables $x$, $y$, $A$. Indeed, we have
\begin{equation}
\label{phi}
\phi=M_{Pl}\frac{1-A}{A},
\end{equation}
\begin{equation}
\label{phit}
{\dot\phi}^2=2xV(\phi)=2xV_0{\left( \frac{1-A}{A}\right) }^m e^{-\lambda\frac{{(1-A)}^n}{A^n}},
\end{equation}
\begin{equation}
\begin{array}{l}
\label{H}
H^2=\frac{{V_{\phi}}^2}{y^2{\dot\phi}^2}=\frac{{V_0}^2}{y^2{\dot\phi}^2}e^{-2\lambda\frac{\phi^n}{{M^n_{Pl}}}}{\left( m\frac{\phi^{m-1}}{{M^m_{Pl}}}-\lambda n\frac{\phi^{m+n-1}}{{M^{m+n}_{Pl}}}\right)}^2 =\\
=\frac{V_0}{2{M^2_{Pl}}xy^2}e^{-\lambda\frac{{(1-A)}^n}{A^n}}{(1-A})^{m-2}A^{2-2n-m}{\left[ mA^n-\lambda n{(1-A)}^n \right]}^2.
\end{array}
\end{equation}
Positive values of the scalar field $\phi$ correspond to $A\in(0; 1)$, whereas negative ones correspond to $A\in(-\infty; 0)\cup(1; +\infty)$ and $A\to\pm\infty$ for $\phi\to-1$.

    Using definitions of $x$, $y$, $A$ (\ref{xyA}) and the expression (\ref{phi}), we derive the following important relations: 
\begin{equation}
\label{VH2}
\frac{V}{2{M_{Pl}}^2H^2}=xy^2A^{2n-2}{\left[ \frac{1-A}{mA^n-\lambda n{(1-A)}^n}\right] }^2,
\end{equation}
\begin{equation}
\begin{array}{l}
\label{HtH2}
\frac{\dot H}{H^2}=-\frac{3}{2}(w+1)+\frac{V}{2{M_{Pl}}^2H^2}(x(w-1)+w+1)=\\
=-\frac{3}{2}(w+1)+xy^2A^{2n-2}(x(w-1)+w+1){\left[ \frac{1-A}{mA^n-\lambda n{(1-A)}^n}\right] }^2,
\end{array}
\end{equation}
\begin{equation}
\label{fttftH}
\frac{\ddot\phi}{\dot\phi H}=-y-3,
\end{equation}
\begin{equation}
\label{wphi}
w_{\phi}=\frac{x-1}{x+1},
\end{equation}
\begin{equation}
\begin{array}{l}
\label{weff}
w_{\rm eff}=\frac{p+p_{\phi}}{\rho+\rho_{\phi}}=-\frac{(2\dot H +3 H^2){M_{Pl}}^2}{3 {M_{Pl}}^2H^2}=-1-\frac{2}{3}\frac{\dot H}{H^2}=\\
=w-\frac{2}{3}xy^2 A^{2n-2}(x(w-1)+w+1){\left[ \frac{1-A}{mA^n-\lambda n{(1-A)}^n}\right] }^2,
\end{array}
\end{equation}
\begin{equation}
\label{rhophi}
\rho_{\phi}=\frac{{\dot\phi}^2}{2}+V(\phi)=V_0(x+1){\left( \frac{1-A}{A}\right) }^m e^{-\lambda\frac{{(1-A)}^n}{A^n}},
\end{equation}
\begin{equation}
\begin{array}{l}
\label{rho}
\rho=3H^2{M_{Pl}}^2-\rho_{\phi}=\\
=\frac{3V_0}{2xy^2}e^{-\lambda\frac{{(1-A)}^n}{A^n}}{(1-A)}^{m-2}A^{2-2n-m}{\left[ m A^n-\lambda n{(1-A)}^n \right]}^2 -V_0(x+1){\left( \frac{1-A}{A}\right) }^m e^{-\lambda\frac{{(1-A)}^n}{ A^n}}=\\
=V_0{\left( \frac{1-A}{A}\right) }^{m-2} e^{-\lambda\frac{{(1-A)}^n}{A^n}}\left\lbrace  \frac{3}{2xy^2A^{2n}}{\left[ m A^n-\lambda n{(1-A)}^n \right]}^2 -(x+1){\left( \frac{1-A}{A}\right) }^2\right\rbrace,
\end{array}
\end{equation}
\begin{equation}
\label{Omegaphi}
\Omega_{\phi}=\frac{\rho_{\phi}}{3{M_{Pl}}^2H^2}=\frac{V}{3{M_{Pl}}^2H^2}(x+1)=\frac{2}{3}(x+1)xy^2A^{2n-2}{\left[ \frac{1-A}{mA^n-\lambda n{(1-A)}^n}\right] }^2.
\end{equation}
Let us note that points of the phase space of the new variables $(x, y, A)$ have the physical sense only if their coordinates $x_0$, $y_0$, $A_0$ give rise to $\rho\geqslant0$ after the substitution into Eq.~(\ref{rho}).

    Let us emphasize that the parameter $\Gamma$ becomes important beyond the standard exponential potential; it governs the dynamics of the underlying system. In general, it provides a yardstick to check for the scaling solutions. For the potential under consideration, we have
\begin{equation}
\label{Gamma}    
\Gamma=1+(n-1)\frac{V}{V_{\phi}\phi}-mn{\left( \frac{V}{V_{\phi}\phi}\right) }^2=1+\frac{(n-1)A^n}{mA^n-\lambda n{(1-A)}^n}-\frac{mnA^{2n}}{{\left[ mA^n-\lambda n{(1-A)}^n\right] }^2},
\end{equation}
where we have used the following relation,
\begin{equation}
\label{VVphi}    
\frac{V}{V_{\phi}\phi}=\frac{A^n}{mA^n-\lambda n{(1-A)}^n}.
\end{equation}
    Let us note that $n=1$, $m=0$ corresponds to the standard case of an exponential potential which gives rise to a scaling solution. It is interesting that $\Gamma\to 1$ for arbitrary values of $n$ and $m$ provided that $A\to 0$ corresponding to $|\phi/M_{Pl}|\to\infty$, which signals the emergence of scaling behavior in the asymptotic regime, while $\Gamma\to\frac{m-1}{m}$ in the other asymptotic limit, $\phi/M_{Pl}\to 0$ (that is, $A\to 1$). In what follows, we shall demonstrate this rigorously.
    
    Taking the derivative of $x$, $y$, $A$ with respect to ($\ln a$), we find the following autonomous system,
\begin{equation}
\label{dx0}
\frac{dx}{d(\ln a)}=-2x(3+y+xy),
\end{equation}
\begin{equation}
\label{dy0}
\frac{dy}{d (\ln a)}=2 x y^2\frac{V_{\phi\phi}V}{{V_{\phi}^2}} -y\frac{\ddot\phi}{\dot\phi H}-y\frac{\dot H}{H^2},
\end{equation}
\begin{equation}
\label{dA0}
\frac{dA}{d (\ln a)}=-2xyA(1-A)\frac{V}{V_{\phi}\phi},
\end{equation}
which after the substitution of the relations (\ref{HtH2}), (\ref{fttftH}), (\ref{Gamma}) and (\ref{VVphi}) has the final form
\begin{equation}
\label{dx}
\frac{dx}{d(\ln a)}=-2x(3+y+xy),
\end{equation}
\begin{equation}
\begin{array}{l}
\label{dy}
\frac{dy}{d (\ln a)}=2 x y^2\left\lbrace   1+\frac{(n-1)A^n}{mA^n-\lambda n{(1-A)}^n}-\frac{mnA^{2n}}{{\left[  mA^n-\lambda n{(1-A)}^n\right]  }^2}\right\rbrace  +y(9/2+y+3w/2)-\\
-xy^3A^{2n-2}(x(w-1)+w+1){\left[ \frac{1-A}{mA^n-\lambda n{(1-A)}^n}\right] }^2,
\end{array}
\end{equation}
\begin{equation}
\label{dA}
\frac{dA}{d (\ln a)}=-\frac{2xyA^{n+1}(1-A)}{mA^n-\lambda n{(1-A)}^n}.
\end{equation}
Once the autonomous system of equations is set up, we can proceed for its analysis, which we do in the following subsections.

\subsection{Stationary points}
    Solving the system (\ref{dx})-(\ref{dA}), we find the stationary points and investigate their stability in the linear approximation {\it \`{a} la} Lyapunov~\cite{nonlinear}. In this case, there exist four fixed points.
\\
\\\textbf{1.} $x=0$, $y=0$, $A\in(-\infty; +\infty)$.
\\This is stationary line. The eigenvalues of the Jacobian matrix associated with the system (\ref{dx})-(\ref{dA}) are given by
\begin{equation}
\begin{array}{l}
L_1=-6<0,\\
L_2=9/2+3w/2>0 \text{~~~~for $w\in[-1; 1]$}.\\
L_3=0
\end{array}
\end{equation}
The eigenvalue $L_2$ is positive for $w\in[-1; 1]$; hence this stationary line is unstable. Using its coordinates $x$,~$y$,~$A$, we calculate the corresponding values of the important quantities using Eqs.~(\ref{HtH2}) and (\ref{fttftH}): $\frac{\dot H}{H^2}=-\frac{3}{2}(w+1)$, $\frac{\ddot\phi}{\dot\phi H}=-3$. Solving them, we obtain $H(t)=\frac{2}{3(w+1)(t-t_0)}$, $\frac{d(\dot\phi)}{\dot\phi}=-3H(t)dt$ and
\begin{equation}
\label{phit1}
\dot\phi(t)={\dot\phi}_1 \left(\frac{t}{t_0}-1\right)^{-\frac{2}{w+1}},
\end{equation}
\begin{equation}
\label{phi1}
\phi(t)=\phi_0+\phi_2\frac{w+1}{w-1}\left(\frac{t}{t_0}-1\right)^{\frac{w-1}{w+1}},
\end{equation}
and the time dependence of the scale factor is
\begin{equation}
\label{a1}
a(t)=a_0\left(\frac{t}{t_0}-1\right)^{\frac{2}{3(w+1)}}.
\end{equation}
Using the continuity equation $\dot\rho+3(w+1)H\rho=0$, we find the time dependence of the energy density of matter,
\begin{equation}
\label{rho1}
\rho(t)=\rho_0\left(\frac{t}{t_0}-1\right)^{-2}.
\end{equation}
Here $t_0$, $\phi_0$, ${\dot\phi}_1$ and $\phi_2$ are constants. $\phi_0$ and $\phi_2$ have the dimension of energy, $t_0$ has the dimension of $\text{(energy)}^{-1}$, $\dot\phi_1$ has the dimension of $(\text{energy})^2$.

    The coordinates of this stationary line do not allow us to reconstruct correctly the behavior of the scalar field $\phi(t)$, and the formula (\ref{phi1}) is approximate. Our numerical investigations (see the next section) confirm the power-law dependence $\dot\phi(t)$ in Eq.~(\ref{phit1}); however, the absolute value of $\dot\phi$ is very small --- close to zero --- and $\phi(t)$ changes very slowly and is equal approximately to the constant $\phi_0$. 

    Substituting the values of the coordinates of this stationary line into formulas (\ref{wphi}) and (\ref{weff}), we obtain $w_{\phi}\to -1$ and $w_{\rm eff}\to w$. As the coordinate $A$ runs along an axis of real numbers, the parameter $\Gamma=\Gamma(A)$ [see Eq.~(\ref{Gamma})] can take various values, and we have $\Gamma=1$ for the point of this line with $A=0$.  
    
     In the particular case of $w=-1$, instead of the solution (\ref{phi1})-(\ref{rho1}), we conclude from Eqs.~(\ref{HtH2}) and (\ref{fttftH}) that the Hubble parameter $H=H_0$, the matter energy density $\rho=\rho_0$, the time derivative of the scalar field decreases as the exponent $\dot\phi(t)={\dot\phi}_1 e^{-3H_0(t-t_0)}$, and the scalar field tends to the constant $\phi(t)=\phi_0-\phi_2 e^{-3H_0(t-t_0)}\to \phi_0$.
\\
\\\textbf{2.} $x=0$, $y=-\frac{3}{2}(w+3)$, $A\in(-\infty; +\infty)$.
\\We find the eigenvalues for this stationary line
 \begin{equation}
\begin{array}{l}
L_1=3+3w\geqslant0 \text{~~~~for $w\in[-1; 1]$},\\
L_2=-9/2-3w/2<0 \text{~~~~for $w\in[-1; 1]$},\\
L_3=0.
\end{array}
\end{equation}
This stationary line is unstable for $w\in(-1; 1]$. Using Eqs.~(\ref{HtH2}) and (\ref{fttftH}), we get $\frac{\dot H}{H^2}=-\frac{3}{2}(w+1)$,
$\frac{\ddot\phi}{\dot\phi H}=\frac{3}{2}(w+1)$. Then 
\begin{equation}
\label{phit2}
\dot\phi(t)={\dot\phi}_1(t/t_0-1),
\end{equation}
\begin{equation}
\label{phi2}
\phi(t)=\phi_0+\phi_2{(t/t_0-1)}^2,
\end{equation}
where $t_0$, $\phi_0$, $\phi_2$ and ${\dot\phi}_1$ are constants. Let us note that the behaviors of the scale factor and the matter energy density are the same as in case of the stationary line \textbf{1} [see Eqs.~(\ref{a1}) and (\ref{rho1})].

    Using the numerical integration, we confirm the time dependences of the scalar field derivative (\ref{phit2}) and the difference $\phi(t)-\phi_0=\phi_2{(t/t_0-1)}^2$. It is found that $|\dot\phi|$ is close to zero and that $\phi(t)\approx \phi_0$.  
    
    Applying formulas (\ref{wphi}) and (\ref{weff}), it is calculated that $w_{\phi}\to -1$ and $w_{\rm eff}\to w$ in this line. The parameter $\Gamma$ is equal~to~$1$ for the point of this stationary line, with the coordinate $A=0$.
    
    For $w=-1$, the obtained solution does not exist, and we find, using Eqs.~(\ref{HtH2}) and (\ref{fttftH}), $H=H_0$, $\rho=\rho_0$, $\dot\phi(t)={\dot\phi}_1$. The numerical integration gives us $\dot\phi(t)=\dot\phi_1\approx 0$, and the scalar field is equal to the constant $\phi(t)\approx\phi_0$.
\\
\\\textbf{3.} $x=\frac{1+w}{1-w}$, $y=\frac{3}{2}(w-1)$, $A=0$.
\\This point corresponds to the scaling solution, and it exists for $w\neq1$ (for realistic fluid, $w=0; 1/3$). We are especially interested in this case; the eigenvalues are
\begin{equation}
\begin{array}{l}
L_1=0,\\
L_{2,3}=3/4\left( w-1\pm\sqrt{(9w+7)(w-1)}\right).
\end{array}
\end{equation}
Real parts of eigenvalues $L_2$, $L_3$ are negative for $w\in(-1; 1)$. As $L_1$ vanishes, one needs an additional check for the determination of the type of stability.

    Since the coordinate $x=\frac{1+w}{1-w}=\frac{{\dot\phi}^2}{2V}\geqslant0$ for $w\in[-1; 1)$, $V(\phi)>0$ and the conditions of existence of the scaling point are
\\\textbf{(1)} $m$ is even, $\forall\phi$,
\\\textbf{(2)} $m$ is odd, $\phi>0$. 

    We substitute coordinates of this point into Eqs.~(\ref{HtH2}) and (\ref{fttftH}) and find that $\frac{\dot H}{H^2}=-\frac{3}{2}(w+1)$, $\frac{\ddot\phi}{\dot\phi H}=-\frac{3}{2}(w+1)$. Therefore, for this point, $\dot\phi(t)={\dot\phi}_1{(\frac{t}{t_0}-1)}^{-1}$, where ${\dot\phi}_1$, $t_0$ are constants, and $a(t)$ and $\rho(t)$ are the same as in Eqs.~(\ref{a1}) and (\ref{rho1}). Actually, the time dependence $\dot\phi(t)$ is more complicated, and we cannot reconstruct it using only fixed point coordinates. 
    
    Substituting $x$, $y$, $A$ of this stationary point into Eqs.~(\ref{wphi}), (\ref{weff}), (\ref{Omegaphi}) and (\ref{Gamma}), we readily check that ~~$w_\phi\to w$, ~~$w_{\rm eff}\to w$, ~~$\Omega_{\phi}\to 0$, ~~$\Gamma\to 1$ as the fixed point is approached.

    The numerical integration of the system (\ref{dx})-(\ref{dA}) (see Figs.~\ref{Fig1}-\ref{Fig3}) shows that the scaling stationary point \textbf{3} is the attractor in phase space $(x, y, A)$ for $(\ln a)\to +\infty$ in some region of the initial data. Therefore, the found scaling regime exists for $t\to +\infty$.
    
    The time behavior of the scalar field $\phi(t)$ cannot be found correctly from the coordinates of this stationary point. Following Ref.~\cite{Sami}, where the potential $V(\phi)=V_0e^{-\lambda\frac{\phi^n}{{M_{Pl}}^n}}$ was considered, we assume for the potential (\ref{pot}) that the time dependence of $\phi(t)$ is given by the following series,
\begin{equation}
\label{phi3}  
\lambda{\left( \frac{\phi}{M_{Pl}}\right)}^n=f_0\ln \left( \frac{t}{t_1}\right) +f_1\ln\left[\ln\left(\frac{t}{t_1}\right)\right] +...~~~~,
\end{equation}  
where $f_0$, $f_1$, $t_1$ are constants. Substituting this form of scaling solution (\ref{phi3}), $a(t)=a_0\left(\frac{t}{t_0}\right)^\frac{2}{3(w+1)}$, into Eq.~(\ref{system3}) and using the fact that $V_{\phi}=V_0\frac{\phi^{m-1}}{{M_{Pl}}^m}e^{-\lambda\frac{\phi^n}{{M_{Pl}}^n}}\left( m-\lambda n\frac{\phi^n}{{M_{Pl}}^n} \right)\to V_0 \frac{\phi^{m-1}}{{M_{Pl}}^m}e^{-\lambda\frac{\phi^n}{{M_{Pl}}^n}}\left(-\lambda n\frac{\phi^n}{{M_{Pl}}^n}\right)$ for $\phi^n\to\pm\infty$, $t\to +\infty$, we obtain
\begin{equation}
\label{eqf0f1t1}
\frac{M_{Pl}{f_0}^{\frac{1}{n}}{\left[\ln \left( \frac{t}{t_1}\right)  \right] }^{\frac{1}{n}-1}}{{\lambda}^{\frac{1}{n}}nt^2}\left( -1+\frac{2}{1+w}\right)- \frac{nV_0 }{M_{Pl}{\lambda}^{\frac{m-1}{n}}}{f_0}^{\frac{m+n-1}{n}}t_1^{f_0}t^{-f_0}{\left[\ln \left(  \frac{t}{t_1}\right)  \right] }^{\frac{m+n-1}{n}-f_1}=0.
\end{equation}
Equating the power-law indices and the coefficients of these two terms, we find the values of $f_0$, $f_1$, $t_1$:
\begin{equation}
\label{f0f1t1}
f_0=2,~~~~ f_1=\frac{2n-2+m}{n},~~~~{t_1}^2=\frac{{M_{Pl}}^2(1-w)}{V_0 n^2(1+w)}2^{\frac{2-n-m}{n}}\lambda^{\frac{m-2}{n}}. 
\end{equation} 

    We can find the character of approach to zero of the quantities $\dot\phi(t)$, $V(\phi(t))$, $\rho_{\phi}(t)$, $\Omega_{\phi}(t)$ in the obtained scaling solution. From Eq.~(\ref{phi3}), it follows that
\begin{equation}
\label{phit3}
\dot\phi(t)=\frac{C_1}{t}{\left[ \ln \left(\frac{t}{t_1}\right)\right] }^{\frac{1}{n}-1}+\frac{C_2}{t}\ln\left[\ln\left(\frac{t}{t_1}\right)\right]{\left[ \ln \left(\frac{t}{t_1}\right) \right]}^{\frac{1}{n}-2}+...~~~~,
\end{equation}
\begin{equation}
\label{V3}
V(\phi(t))=\frac{C_3}{t^2}{\left[ \ln \left(\frac{t}{t_1}\right)\right] }^{\frac{2}{n}-2}+\frac{C_4}{t^2}\ln\left[\ln\left(\frac{t}{t_1}\right)\right]{\left[ \ln \left(\frac{t}{t_1}\right)\right]}^{\frac{2}{n}-3}+...~~~~,
\end{equation}
where $C_1=\frac{M_{Pl}}{n}{\left( \frac{2}{\lambda}\right) }^{\frac{1}{n}}$, $C_2=\frac{M_{Pl}(1-n)(m-2+2 n)}{2 n^3}{\left( \frac{2}{\lambda}\right) }^{\frac{1}{n}}$, $C_3=\frac{{M_{Pl}}^2(1-w)}{2 n^2(1+w)}{\left( \frac{2}{\lambda}\right) }^{\frac{2}{n}}$, $C_4=\frac{{M_{Pl}}^2m(m-2+2 n)(1-w)}{4 n^4(1+w)}{\left( \frac{2}{\lambda}\right) }^{\frac{2}{n}}$ are constants. Then 
\begin{equation}
\begin{array}{l}
\label{rhophi3}
\rho_{\phi}(t)=\frac{{\dot\phi}^2}{2}+V(\phi)=\frac{C_1^2/2+C_3}{t^2}{\left[ \ln \left(\frac{t}{t_1}\right)\right] }^{\frac{2}{n}-2}+\frac{C_1C_2+C_4}{t^2}\ln\left[\ln\left(\frac{t}{t_1}\right)\right]{\left[ \ln \left(\frac{t}{t_1}\right)\right]}^{\frac{2}{n}-3}+...~~~~.
\end{array}
\end{equation} 
Therefore, we find that $\rho_{\phi}$ tends to zero for $n>1$, $t\to+\infty$ faster than $\rho(t)=\rho_0 \left(\frac{t}{t_0}\right)^{-2}$ in the scaling regime. Keeping the first-order term in Eq.~(\ref{rhophi3}), we have
\begin{equation}
\begin{array}{c}
\label{rhophirho3}
\frac{\rho_{\phi}}{\rho}\to\frac{{C_1}^2/2+C_3}{\rho_0 {t_0}^2}{\left[ \ln \left(\frac{t}{t_1}\right)\right]}^{\frac{2(1-n)}{n}}=\frac{{M_{Pl}}^2}{n^2(1+w)\rho_0 {t_0}^2}{\left( \frac{2}{\lambda}\right)}^{\frac{2}{n}}{\left[ \ln \left(\frac{t}{t_1}\right)\right]}^{\frac{2(1-n)}{n}} \propto {\left[ \ln \left(\frac{t}{t_1}\right)\right]}^{\frac{2(1-n)}{n}} \propto {\left[ \ln \left(\frac{a}{a_0}\right)\right]}^{\frac{2(1-n)}{n}}\to 0,\\ 
\\\text{for ~~$n>1$,~~ $w\neq-1$,~~ $t\to+\infty$}.
\end{array}
\end{equation} 
The quantity $\Omega_{\phi}$ has the following asymptotic behavior,
\begin{equation}
\begin{array}{c}
\label{Omegaphi3}
\Omega_{\phi}(t)=\frac{\rho_{\phi}}{3{M_{Pl}}^2H^2}\to\frac{3(w+1)}{4n^2}{\left( \frac{2}{\lambda}\right)}^{\frac{2}{n}}{\left[ \ln \left(\frac{t}{t_1}\right)\right]}^{\frac{2(1-n)}{n}} \propto {\left[ \ln \left(\frac{t}{t_1}\right)\right]}^{\frac{2(1-n)}{n}} \propto {\left[ \ln \left(\frac{a}{a_0}\right)\right]}^{\frac{2(1-n)}{n}}\to 0,\\ 
\\\text{for ~~$n>1$,~~ $w\neq-1$,~~ $t\to+\infty$},
\end{array}
\end{equation}    
where the Hubble parameter in the scaling solution $H(t)=\frac{2}{3(w+1)t}$ has been substituted. It is seen that $\Omega_{\phi}\to \frac{3(w+1)}{{\lambda}^2}=const$, for $n=1$. The same behavior of $\Omega_{\phi}$ is obtained in the scaling regime for the model with the potential $V(\phi)=V_0 {\rm exp}(-\lambda\frac{\phi^n}{{M_{Pl}}^n})$ in Ref.~\cite{Sami}. 

    Now we obtain the character of tendency $w_{\phi}\to w$, using Eqs.~(\ref{phit3}) and (\ref{V3}) and keeping only the first dominated two terms in them:
\begin{equation}
\begin{array}{l}
\label{wphieps1}
w_{\phi}(t)=\frac{{\dot\phi}^2/2-V(\phi)}{{\dot\phi}^2/2+V(\phi)}\approx\frac{(C_1^2/2-C_3){\left[ \ln \left(\frac{t}{t_1}\right)\right] }^{\frac{2}{n}-2}+(C_1C_2-C_4)\ln\left[\ln\left(\frac{t}{t_1}\right)\right]{\left[ \ln \left(\frac{t}{t_1}\right)\right]}^{\frac{2}{n}-3}}{(C_1^2/2+C_3){\left[ \ln \left(\frac{t}{t_1}\right)\right] }^{\frac{2}{n}-2}+(C_1C_2+C_4)\ln\left[\ln\left(\frac{t}{t_1}\right)\right]{\left[ \ln \left(\frac{t}{t_1}\right)\right]}^{\frac{2}{n}-3}}=\\
\\=\frac{C_1^2/2-C_3+(C_1C_2-C_4)\ln\left[\ln\left(\frac{t}{t_1}\right)\right]/\ln \left(\frac{t}{t_1}\right)}{C_1^2/2+C_3+(C_1C_2+C_4)\ln\left[\ln\left(\frac{t}{t_1}\right)\right]/\ln \left(\frac{t}{t_1}\right)}.
\end{array}
\end{equation}
Denoting $\varepsilon(t)=\ln\left[\ln\left(\frac{t}{t_1}\right)\right]/\ln \left(\frac{t}{t_1}\right)$, which tends to zero for $t\to+\infty$, and taking into account $\frac{1}{1+x}\approx 1-x$ for $|x|<1$, we find
\begin{equation}
\begin{array}{l}
\label{wphieps2}
w_{\phi}(t)\approx\frac{C_1^2/2-C_3+(C_1C_2-C_4)\varepsilon(t)}{C_1^2/2+C_3+(C_1C_2+C_4)\varepsilon(t)}\approx\frac{{C_1}^2/2-C_3+(C_1C_2-C_4)\varepsilon(t)}{{({C_1}^2/2+C_3)}^2}\Big{[} C_1^2/2+C_3-(C_1C_2+C_4)\varepsilon(t)\Big{]}=\\
\\=\frac{{C_1}^2/2-C_3}{{C_1}^2/2+C_3}+\frac{(C_1C_2-C_4)({C_1}^2/2+C_3)-(C_1C_2+C_4)({C_1}^2/2-C_3)}{{({C_1}^2/2+C_3)}^2}\varepsilon(t)+\frac{{C_4}^2-{C_1}^2{C_2}^2}{{({C_1}^2/2+C_3)}^2}\varepsilon^2(t).
\end{array}
\end{equation}
Neglecting the term with $\varepsilon^2(t)$ and calculating $\frac{{C_1}^2/2-C_3}{{C_1}^2/2+C_3}=w$,~~  $(C_1C_2-C_4)({C_1}^2/2+C_3)-(C_1C_2+C_4)({C_1}^2/2-C_3)=C_1(2C_2C_3-C_1C_4)$,~~ $\frac{C_1(2C_2C_3-C_1C_4)}{{({C_1}^2/2+C_3)}^2}=\frac{(w^2-1){(m-2+2 n)}^2}{4 n^2}$, we finally obtain
\begin{equation}
\begin{array}{c}
\label{wphieps3}
w_{\phi}(t)\approx w+(w^2-1){\Big{[}(m-2+2 n)/(2 n)\Big{]}}^2\frac{\ln\left[\ln\left(\frac{t}{t_1}\right)\right]}{\ln \left(\frac{t}{t_1}\right)}\to w, ~~\text{for $t\to+\infty$}.
\end{array}
\end{equation}
The time dependent correction to $w$ in Eq.~(\ref{wphieps3}) is nonvanishing for generic values of $m$, $n$ and $w$ ($n>1$, $m\geqslant 0$, and $w \neq \pm 1$). Also, in the asymptotic regime, the time dependent correction to $w$ proportional to $\ln\big{[}\ln(t)\big{]}/\ln(t)$ decays uniformly, and the dependence on exponents $n$ and $m$ enters into the proportionality constant only. The latter is a manifestation of the fact that $\Gamma \to 1$ in the asymptotic regime irrespective of the numerical values of the exponents. Secondly, Eq.~(\ref{wphieps3}) gives the same asymptotic expression for $w_{\phi}$ as the model with $m=0$ found in Ref.~\cite{Sami}, which reduces to the standard result ($w_{\phi}=w$) for $n=1$. 

    The preceding analysis completes our description of the scaling fixed point. Let us note, that for $w=-1$, coordinates $x$, $y$ of this fixed point and the stationary line \textbf{2} coincide, and we have the same behavior of cosmological quantities, $H=H_0$, $\rho=\rho_0$, $\dot\phi(t)={\dot\phi}_1\approx 0$, $\phi(t)\approx\phi_0$, where $\phi_1\approx 0$ is found by the numerical investigations. We next come to the description of the last fixed point of the dynamical system under consideration.
\\
\\\textbf{4.} $x=-\frac{m(1+w)}{wm-m+4}$, $y=\frac{3(wm-m+4)}{2(m-2)}$, $A=1$.
\\This stationary point exists for $m\neq0$, $m\neq2$, $w\neq\frac{m-4}{m}$. The eigenvalues are found as
\begin{equation}
\begin{array}{l}
L_1=-\frac{3(1+w)}{m-2}<0 \text{~~~~for $m>2$, $w\in(-1; 1]$,}\\
L_{2,3}=\frac{3}{4(m-2)}\left( f_1(m,w)\pm\sqrt{f_2(m,w)}\right) \text{~~~~$Re(L_{2,3})<0$~~ for $m>2$, $w\in\left(-1; \frac{m-6}{m+2}\right)$,} 
\end{array}
\end{equation}
where $f_1(m,w)=w(m+2)-m+6$,
\\\text{~~~~~~~~}$f_2(m,w)=(9m^2-12m+4)w^2+2(20m-m^2-20)w-7m^2+36m-28$.
    
    We note that the function $f_1(m,w)<0$~~ for ~~$w<\frac{m-6}{m+2}$, ~~while ~~$f_2(m,w)<0$~~ for ~~$w_1<w<w_2$
 \\where $w_{1,2}=\frac{20+m^2-20m\mp 8\sqrt{(m-1){(m-2)}^3}}{{(3m-2)}^2}$.
\\Thus the imaginary parts of the eigenvalues $L_{2,3}$ appear for $w_1 < w < w_2$.  

    We find that $w_1<\frac{m-6}{m+2}<w_2$. Moreover, when $w\leqslant w_1$, the eigenvalues $L_3<L_2<0$; when $w\geqslant w_2$, the eigenvalue $L_2>0$. Therefore, taking into account that $L_1<0$ for $m>2$, $w\in(-1; 1]$, we obtain the stability conditions of point~\textbf{4}: it is a stable node for $w\in(-1; w_1]$, $m>2$, and a stable focus for $w\in\left(w_1;\frac{m-6}{m+2}\right)$, $m>2$. 
  
    For example, for a ordinary matter $w=0$, the stable point exits when $m>6$, and for radiation, $w=\frac{1}{3}$ the stable point exists when $m>10$. For $2<m<6$, this point behaves like an unstable (saddle) one for both radiation and matter. For the case $m=6$, with $w=0$ it behaves like a center and for $w=\frac{1}{3}$ it behaves like a saddle point. 

    The coordinate $x$ of this fixed point can be either positive or negative depending on $m$, $w$. Taking into account the definition of $x=\frac{{\dot\phi}^2}{2V}$ we find the conditions of existence of this point:
\\\textbf{(1)} $m$ is even, $\forall\phi$, or $m$ is odd, $\phi>0$, and $w\in\Big{[}-1; \frac{m-4}{m}\Big{)}$,
\\\textbf{(2)} $m$ is odd, $\phi<0$, and $w\in\Big{(}\frac{m-4}{m}; 1\Big{]}$.
\\
\\Since $\frac{m-4}{m} > \frac{m-6}{m+2}$ for $m>2$, for odd $m$ and $\phi<0$, the point \textbf{4} is unstable.

     Time behavior of the scale factor, the scalar field, and the energy density of matter are calculated in the same way as for the previous fixed points. Since for this point $\frac{\dot H}{H^2}=-\frac{3}{2}(w+1)$, $\frac{\ddot\phi}{\dot\phi H}=\frac{3m(1+w)}{2(2-m)}$, we find the following solution around this stationary point $\dot\phi(t)={\dot\phi}_1{\left(\frac{t}{t_0}-1\right)}^{\frac{m}{2-m}}$ and
\begin{equation}
\label{phi4}
\phi(t)=\phi_0{\left(\frac{t}{t_0}-1\right)}^{\frac{2}{2-m}},
\end{equation}
where $t_0$, ${\dot\phi}_1$, $\phi_0$ are constants. The time dependences of the scale factor and the energy density of matter coincide with those in stationary line \textbf{1} [Eqs.~(\ref{a1}) and (\ref{rho1})]. As the fixed point coordinate $A\to 1$, the scalar field $\phi$ tends to zero in the found asymptotic solution. Therefore, it exists for ~~$0<m<2$, $t\to t_0$~~ or ~~$m>2$, $t\to+\infty$.~~ The constant $\phi_0$ is found after the substitution of the obtained solution into Eq.~(\ref{system3}),
\begin{equation}
\label{eqphi0}
\frac{2m \phi_0}{{(2-m)}^2 {t_0}^2}\left(\frac{t}{t_0}-1\right)^{\frac{2(m-1)}{2-m}}+\frac{4 \phi_0}{(2-m)(1+w) {t_0}^2}\left(\frac{t}{t_0}-1\right)^{\frac{2(m-1)}{2-m}}+V_0 m\frac{{\phi_0}^{m-1}}{{M_{Pl}}^m}\left(\frac{t}{t_0}-1\right)^{\frac{2(m-1)}{2-m}}=0,
\end{equation}
where we have used $V_{\phi}=V_0 \frac{\phi^{m-1}}{{M_{Pl}}^m}e^{-\lambda\frac{\phi^n}{{M_{Pl}}^n}}\left( m-\lambda n\frac{\phi^n}{{M_{Pl}}^n}\right)\to V_0 m \frac{\phi^{m-1}}{{M_{Pl}}^m}$ for $\phi \to 0$. From Eq.~(\ref{eqphi0}), it follows that
\begin{equation}
\label{phi0}
{\phi_0}^{m-2}=\frac{2{M_{Pl}}^m(m(1-w)-4)}{{t_0}^2 V_0 m (1+w) {(2-m)}^2}. 
\end{equation} 
    Using values of coordinates of this stationary point, it is obtained that ~~$\Gamma\to\frac{m-1}{m}$, ~~$w_{\phi}\to~\frac{w m+2}{m-2}$, ~~$w_{\rm eff}\to w$. In the discussion to follow, we shall numerically confirm the aforesaid results based upon linear approach.

    For $w=-1$, we have the same situation as in the previous point: $H=H_0$, $\rho=\rho_0$, $\dot\phi(t)=\dot\phi_1\approx 0$, $\phi(t)\approx\phi_0$, where $\phi_1\approx 0$ is obtained numerically.

\subsection{Case of exponential potential: Standard autonomous system versus our framework}
\label{secc}
    In this subsection, we shall demonstrate how our dynamical system  explicitly captures the results corresponding to the standard exponential potential discussed in the literature \cite{Copeland,Barreiro:1999zs}. The latter is associated with a special case obtained by setting $m=0$ and $n=1$ in expression (\ref{pot}). For the sake of comparison, let us briefly mention the standard definition of autonomous variables and the corresponding autonomous system\footnote{The variable $\lambda_s$ is referred to the ``slope'' of the potential, defined such that $\lambda_s=\lambda$ for exponential potential $V\propto {\rm exp}(-\lambda \phi/M_{Pl})$.},
\begin{equation}
\label{anormal}
X\equiv\frac{\dot{\phi}}{\sqrt{6}HM_{Pl}};~~~~
Y\equiv\frac{\sqrt{V}}{\sqrt{3} H};~~~~\lambda_s\equiv 
-M_{Pl}\frac{ V_\phi}{V};~~~~\Gamma\equiv\frac{V V_{\phi \phi}}{{V_{\phi}}^2}.
\end{equation}
The autonomous system then has following symbolic form,
\begin{equation}
\label{snormal}
\frac{dX}{d(\ln a)}=F(X,Y,\lambda_s);~~~~\frac{dY}{d(\ln a)} =G(X,Y,\lambda_s);~~~~\frac{d\lambda_s}{d(\ln a)} =-\sqrt{6}\lambda^2_s(\Gamma-1)x  
\end{equation}
where $F$ and $G$ are functions\footnote{The concrete forms of functions $F$ and $G$ are not relevant to our discussion.} of $X$, $Y$ and $\lambda_s$. Physical quantities are then expressed through these variables --- for instance, $\Omega_{\phi}=X^2+Y^2$ and $w_{\rm eff}=w+(1-w) X^2-(1+w)Y^2$. Let us note that, for the standard exponential potential, $V\propto {\rm exp}(-\lambda \phi/M_{Pl})$, the last equation in Eq.~(\ref{snormal}) becomes redundant; it has trivial information --- namely, the slope of the exponential potential is constant. The autonomous variables defined in Eq.~(\ref{anormal}) are convenient ones for the search of scaling as well as for field dominated fixed points. However, this formalism is not suitable for investigation of scaling solutions in the asymptotic regime ($|\phi|\to\infty$), as the autonomous variables (\ref{anormal}) do not explicitly contain the dependence upon $\phi$. Thus while dealing with a general class of potential such as Eq.~(\ref{pot}), one needs to invent other variables that could capture the said feature. Let us also point out that, apart from the scaling solution, the autonomous system with exponential potential (\ref{snormal}) exhibits one more stable fixed point that corresponds to the field dominant solution --- namely,~ $\Omega_{\phi}=1$;~ $w_{\rm eff}=w_{\phi}=-1+\lambda^2/3$ --- which is stable provided that $\lambda^2<3(1+w)$. This solution can give rise to late-time acceleration provided that $\lambda^2<2$.

    As already mentioned, the autonomous system (\ref{snormal}) is not convenient for the investigation of scaling solutions in the asymptotic regime. We have demonstrated that our system (\ref{dx})-(\ref{dA}) is adequate for said purpose. However, the field dominant fixed point is not visible in our framework, though it is very much there\footnote{For $m \geqslant 0$, $n>1$, there exists no field dominant solution; thus it not surprising that, in our general analysis, this fixed point for $m=0$, $n=1$ is not seen.}. To this effect, let us specialize to $m=0$, $n=1$. In that case, the dynamical system under consideration assumes the following form in our framework,
\begin{eqnarray}
\label{dyneqn2}
\frac{dx}{d(\ln a)}&=&-2x(3+y+xy)\label{dx2}\\
\frac{dy}{d (\ln a)}&=&y \left[ 2xy+9/2+y+3w/2-\frac{x y^2}{\lambda^2} (x(w-1)+w+1)\right] \label{dy2}\\
\frac{dA}{d (\ln a)}&=&\frac{2}{\lambda} A^2 x y\label{dA2},
\end{eqnarray}
where the autonomous variables $x$, $y$ and $A$ are given by Eq.~(\ref{xyA}).
\\
In this case, $\Omega_{\phi}$ and $w_{\rm eff}$ acquire the following form [see Eqs.~(\ref{Omegaphi}) and (\ref{weff})]:
\begin{equation}
\label{Omegaphiweff}
\Omega_{\phi}=\frac{2}{3\lambda^2}(x+1)xy^2,~~~~ w_{\rm eff}=w-\frac{2}{3\lambda^2}xy^2 (x(w-1)+w+1).
\end{equation}

    Clearly, the variable $A$ drops out from Eqs.~(\ref{dx2}) and (\ref{dy2}) in the case of $m=0$, $n=1$ and equations for $x$ and $y$ can be solved without reference of the third equation, which thereby becomes redundant. The latter demonstrate the consistency of our autonomous system. Solving Eqs.~(\ref{dx2}) and (\ref{dy2}), we obtain the fixed point: 
\\ 
\\\textbf{1a.} $x=\frac{\lambda^2}{6-\lambda^2}$,~~ $y=\frac{1}{2}\left( \lambda^2-6\right)$.
\\ 
Analysis shows that \textbf{1a} is an attractor for $\lambda^2<3(1+w)$ (see Appendix C); for this fixed point, $\Omega_{\phi}=1$, and $w_{\rm eff}=w_\phi=-1+\frac{\lambda^2}{3}$, which is certainly the field dominant solution\footnote{$w_{\rm eff}=\Omega_{\phi} w_{\phi}+\Omega_m w$, where $\Omega_m$ stands for the background matter with the equation of state parameter $w$ such that $\Omega_{\phi}+\Omega_m=1$. Obviously for the field dominant solution ($\Omega_{\phi}=1$),~~ $w_{\rm eff}=w_{\phi}$.} with $a(t)=a_0{\left( \frac{t}{t_0}\right) }^{\frac{2}{\lambda^2}}\to+\infty$,~~ $\phi(t)=\phi_0\ln\left( \frac{t}{t_0}\right)\to\pm\infty$,~~ $t\to+\infty$ as mentioned above \cite{Copeland,Barreiro:1999zs}. 
\\

    It is also easy to check for the second fixed point corresponding to the standard scaling solution in our formalism. Indeed, we obtain the following:
\\
\\\textbf{2a.} $x=\frac{1+w}{1-w}$,~~ $y=\frac{3(w-1)}{2}$.
\\ 
In this case, $\Omega_{\phi}=\frac{3 (w+1)}{\lambda^2}$, $w_{\rm eff}=w_{\phi}=w$, the fixed point is stable for $\lambda^2>3(1+w)$, \textit{\`{a} la} the standard scaling solution with $a(t)=a_0{\left( \frac{t}{t_0}\right) }^{\frac{2}{3(w+1)}}\to+\infty$,~~ $\phi(t)=\phi_0 \ln\left( \frac{t}{t_0}\right)\to\pm\infty$,~~ $t\to+\infty$ \cite{Copeland,Barreiro:1999zs}. Using the asymptotic expressions, we demonstrated above, in a complicated manner, that point \textbf{3} in our analysis, which corresponds to the asymptotic scaling solution, gives rise to the fixed point \textbf{2a} for $m=0$, $n=1$; see Eqs.~(\ref{Omegaphi3}) and (\ref{wphieps3}).
 
    The aforesaid includes the complete information about cosmological dynamics of scalar field with exponential potential. Let us comment on the third equation (\ref{dA2}) which, at the onset, does not look trivial. For both the fixed points {\bf 1a} and {\bf 2a}, $H(t)\propto 1/t$, Eq.~(\ref{dA2}) then reduces to $\dot{\phi}\propto 1/t$ or $\phi(t) \propto \ln(t)$. This information is already contained in (\ref{dx2}) and (\ref{dy2}), hence (\ref{dA2}) is redundant though not trivial. 
 
    In this subsection, by specializing our autonomous system (\ref{dx})-(\ref{dA}) to $m=0$, $n=1$, we have explicitly shown the existence and the stability of the field dominant solution. This solution might play an important role in the framework of quintessential inflation if the complicated potential is capable of giving rise to successful inflation, and reconciling with the nucleosynthesis constraint could mimic $V\propto {\rm exp}(-\lambda \phi/M_{Pl})$,~ ($\lambda^2<2$)-like behavior at late stages \cite{Dimopoulos:2017zvq}; see the Appendix C for details.

\section{Numerical investigations}
    In the preceding section, we found the fixed points of the dynamical system under consideration and discussed their stability. In what follows, we shall confirm the details using numerical integration of equations of motion. We integrate numerically two first-order systems of differential equations for several sets of fixed parameters, $n>1$, $m\geqslant0$, $\lambda>0$, $V_0>0$: 
\\\textbf{(1)} The system with initial variables $\phi$, $\Phi=\dot\phi$, $H$, which is derived from Eqs.~(\ref{system1})-(\ref{system3}),
\begin{equation}
\label{system11}
\frac{d\phi}{d (\ln a)}=\frac{\Phi}{H},
\end{equation}
\begin{equation}
\label{system21}
\frac{d\Phi}{d (\ln a)}=-3\Phi-\frac{V_0 e^{-\lambda\frac{\phi^n}{{M_{Pl}}^n}}}{{M_{Pl}}^m H}\left( m\phi^{m-1}-\lambda n\frac{\phi^{m+n-1}}{{M_{Pl}}^n}\right),
\end{equation}
\begin{equation}
\label{system31}
\frac{d H}{d (\ln a)}=-\frac{1}{2{M_{Pl}}^2H}\left\lbrace \Phi^2+(1+w)\left[ 3{M^2_{Pl}}H^2-\frac{\Phi^2}{2}-V_0{\left( \frac{\phi}{M_{Pl}}\right) }^m e^{-\lambda\frac{\phi^n}{{M_{Pl}}^n}}\right] \right\rbrace ,
\end{equation}    
where $\frac{d}{d (\ln a)}\equiv\frac{d}{H dt}$ and $\rho=3{M^2_{Pl}}H^2-\frac{\Phi^2}{2}-
V_0{\left( \frac{\phi}{M_{Pl}}\right) }^m e^{-\lambda\frac{\phi^n}{{M_{Pl}}^n}}$ have been substituted in.
\\
\\\textbf{(2)} The autonomous system with dimensionless variables $x$, $y$, $A$ [Eqs.~(\ref{dx})-(\ref{dA})]. 

\begin{figure}[hbtp]
\includegraphics[scale=0.46]{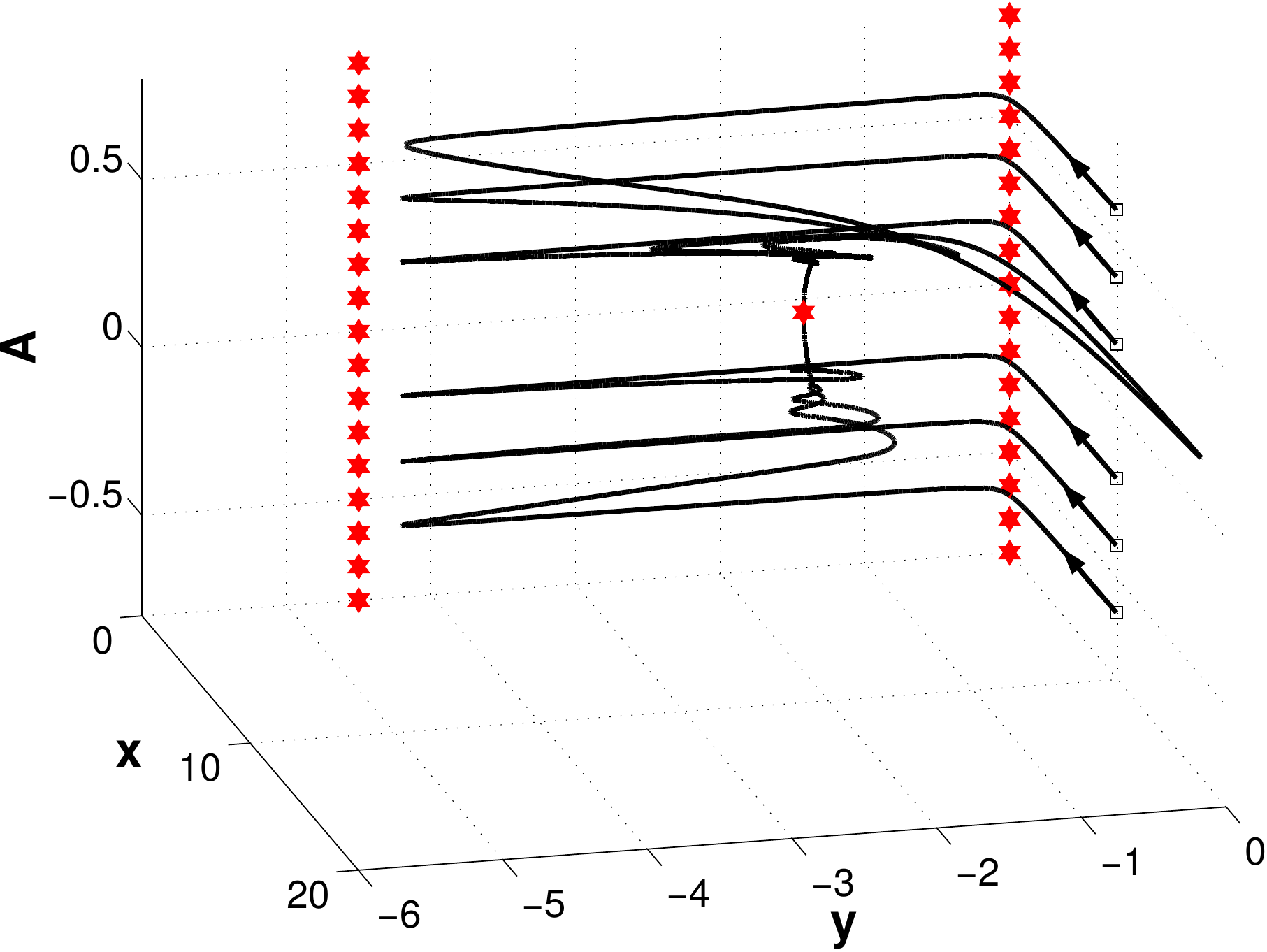}\qquad
\includegraphics[scale=0.46]{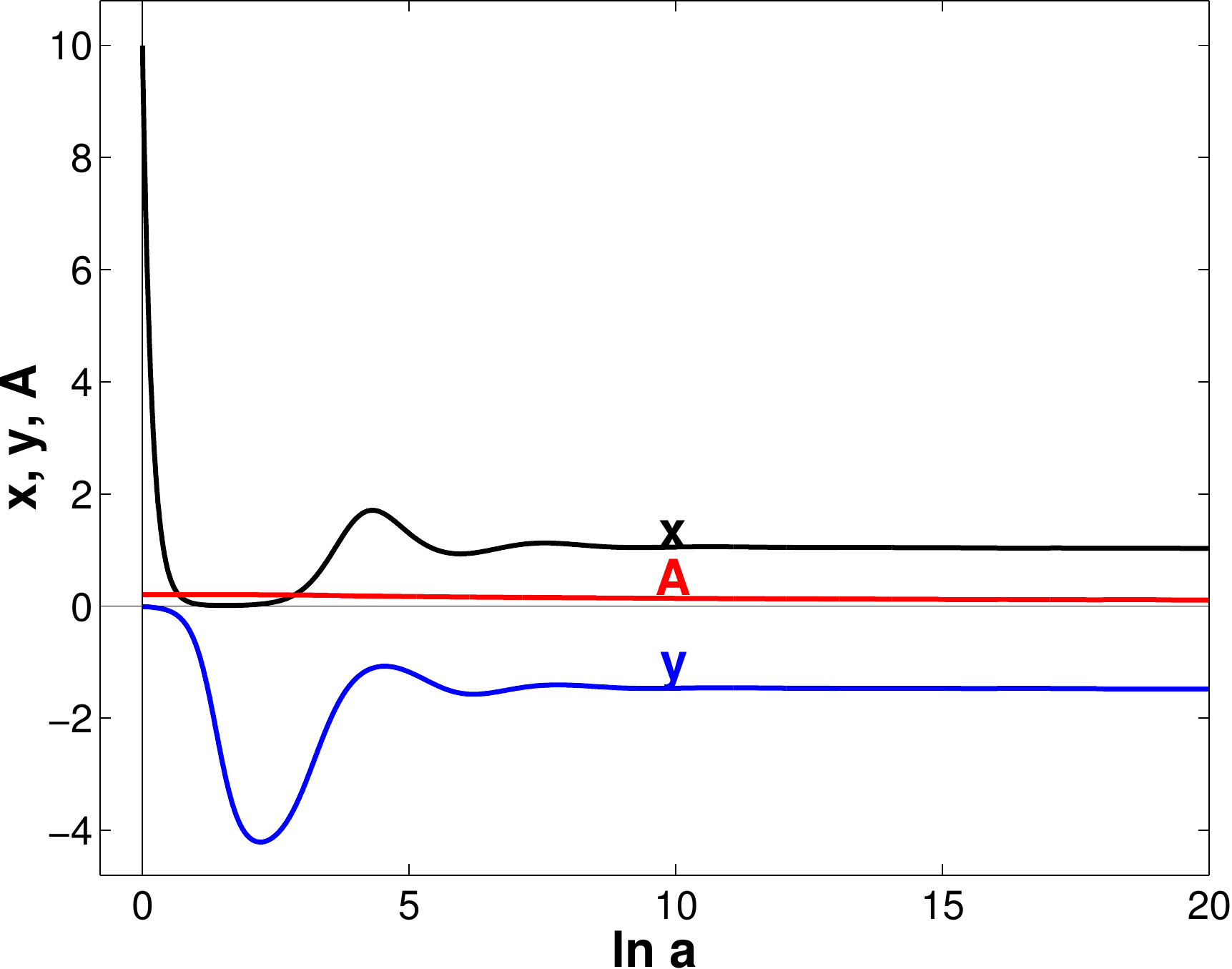}
\caption{(Left panel) Several trajectories in the phase space $(x, y, A)$~~ and ~~(right panel) the evolution of variables $x$,~$y$,~$A$. In the left plot, vertical straight lines of red stars are stationary lines \textbf{1}, \textbf{2}. The separate red star is the scaling solution point \textbf{3}. Squares  denote initial data, which are chosen as ~~$x(0)=10$,~~ $y(0)=-0.01$,~~ $A(0)$~~ from ~~$-0.6$ to $-0.2$~~ with step ~~$0.2$~~ and from ~~$0.2$ to $0.6$~~ with step $0.2$. For the right graph starting values are $\phi(0)=4$,~~ $\dot\phi(0)=0.0015$,~~ $H(0)=0.06$.~~ The parameters are ~~$n=2$, ~~$m=0$, ~~$V_0=1$,~~ $\lambda=1$,~~ $w=0$, ~~${M_{Pl}}^2=1$.}
\label{Fig3}
\end{figure}

\begin{figure}[hbtp]
\includegraphics[scale=0.46]{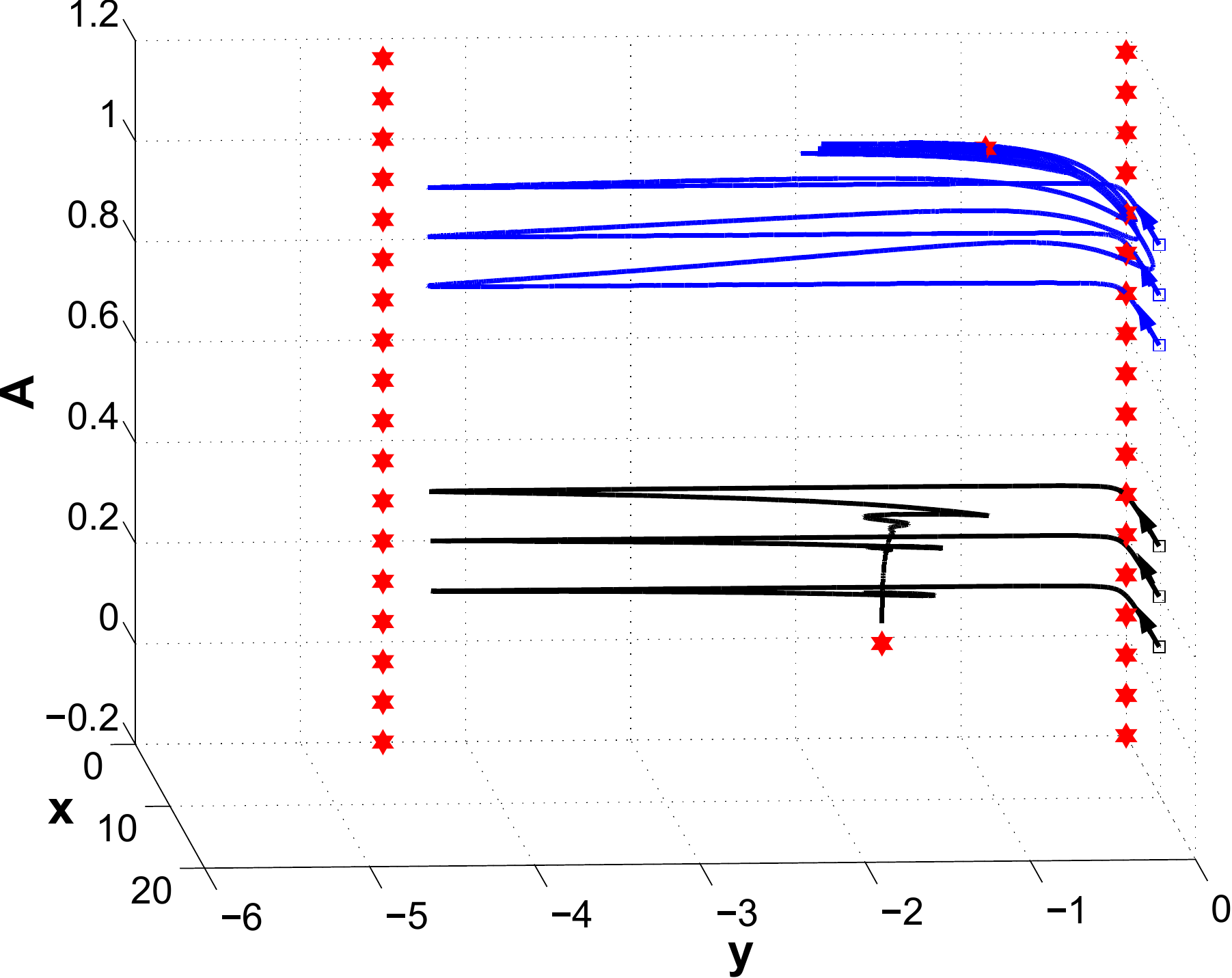}\qquad
\includegraphics[scale=0.46]{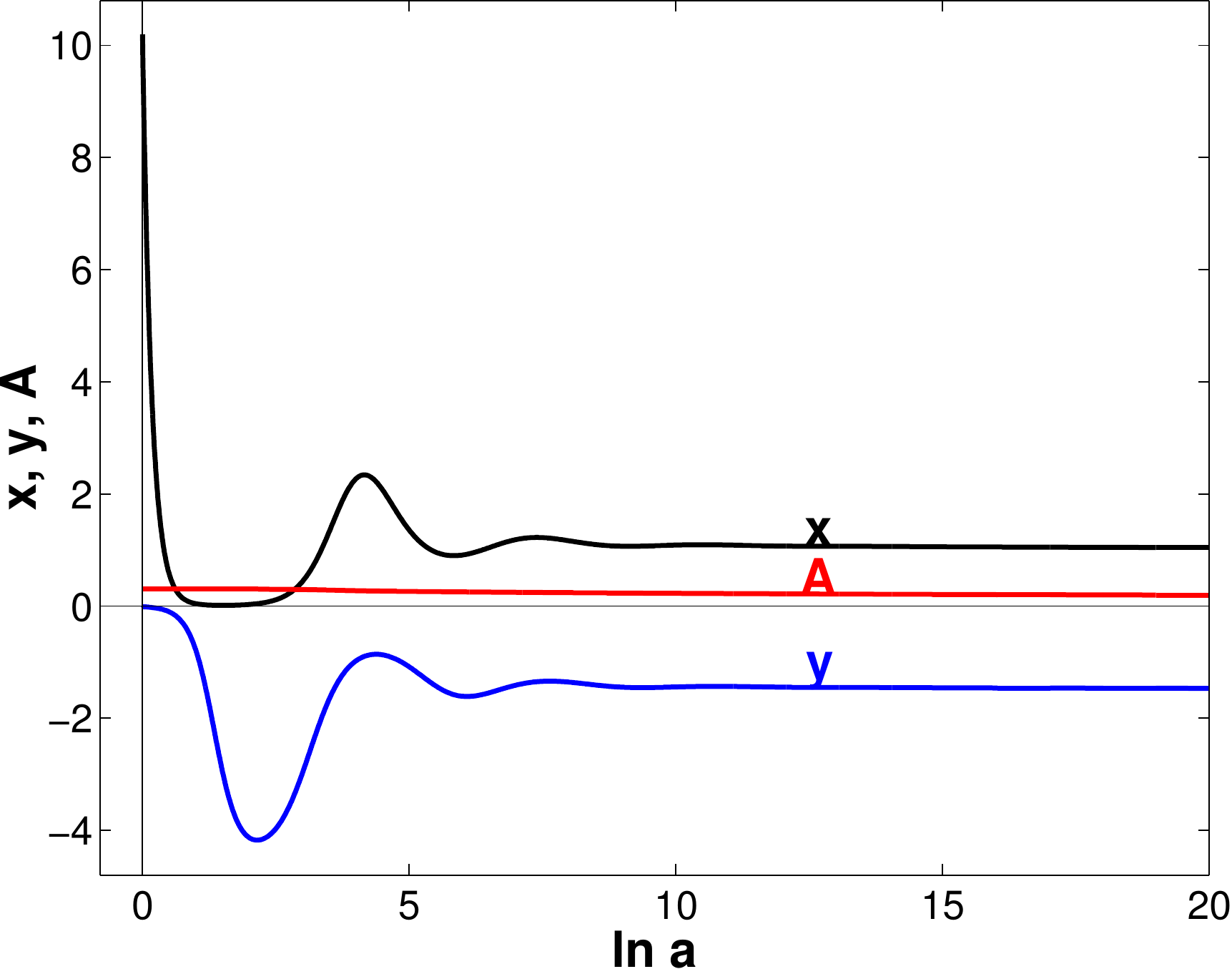}
\caption{(Left panel) Several trajectories in the phase space $(x, y, A)$~~ and ~~(right panel) the evolution of variables $x$,~$y$,~$A$. In the left graph, vertical straight lines of red stars are stationary lines \textbf{1}, \textbf{2}. The separate red stars are the scaling solution point \textbf{3} and the point \textbf{4}. Squares denote initial data, which are chosen as ~~$x(0)=10$,~~ $y(0)=-0.01$,~~ $A(0)$~~ from ~~$0.1$ to $0.3$~~ with step ~~$0.1$~~ for the black curves and ~~$x(0)=10$,~~ $y(0)=-0.01$,~~ $A(0)$ from ~~$0.7$ to $0.9$~~ with step ~~$0.1$~~ for the blue curves. For the right plot, the starting values are $\phi(0)=2.3$,~~ $\dot\phi(0)=0.19$,~~ $H(0)=10$.~~ The parameters are ~~$n=3$, ~~$m=7$, ~~$V_0=1$,~~ $\lambda=1$, ~~$w=0$, ~~${M_{Pl}}^2=1$.}
\label{Fig4}
\end{figure}

\begin{figure}[hbtp]
\includegraphics[scale=0.46]{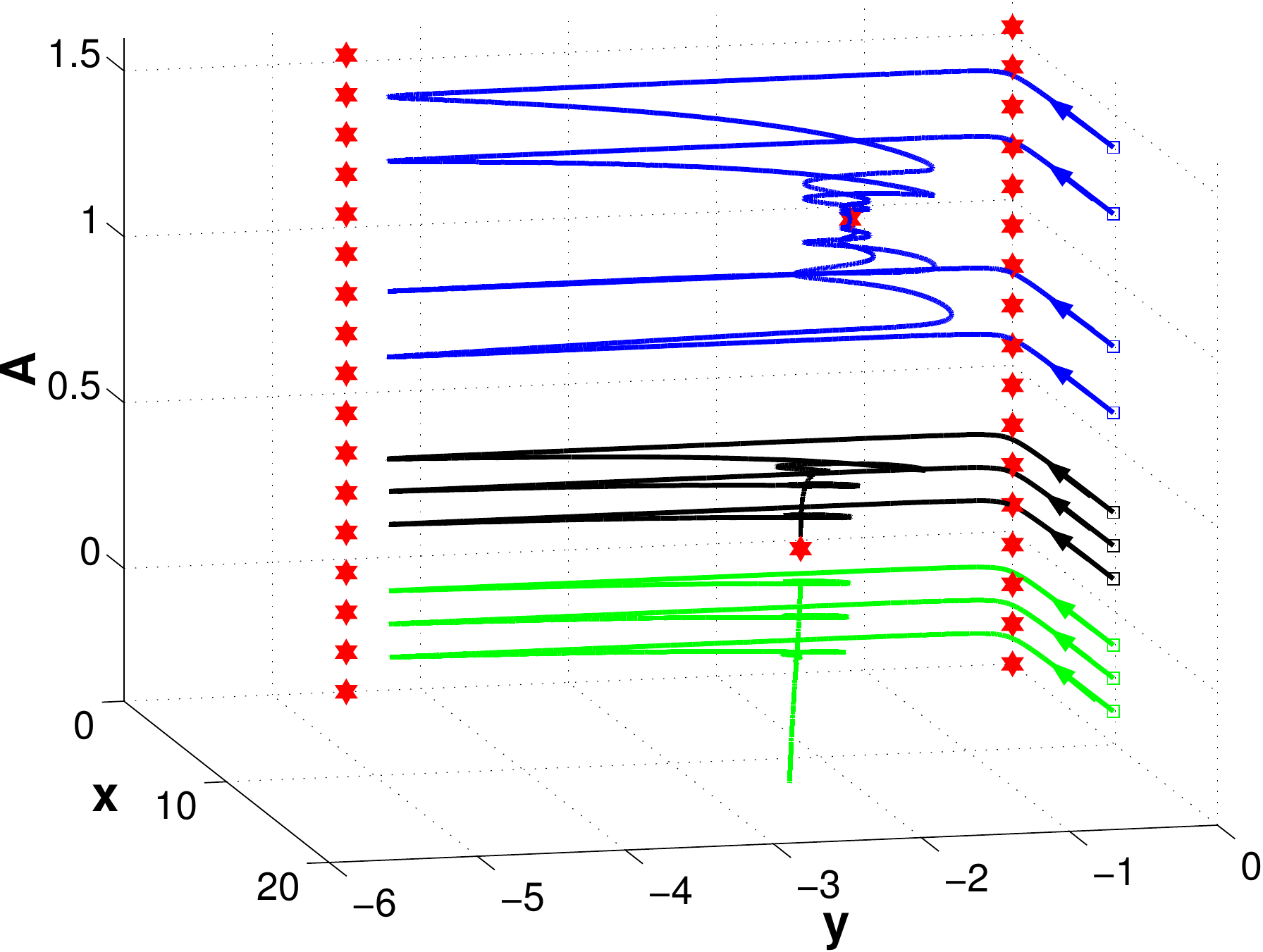}\qquad
\includegraphics[scale=0.46]{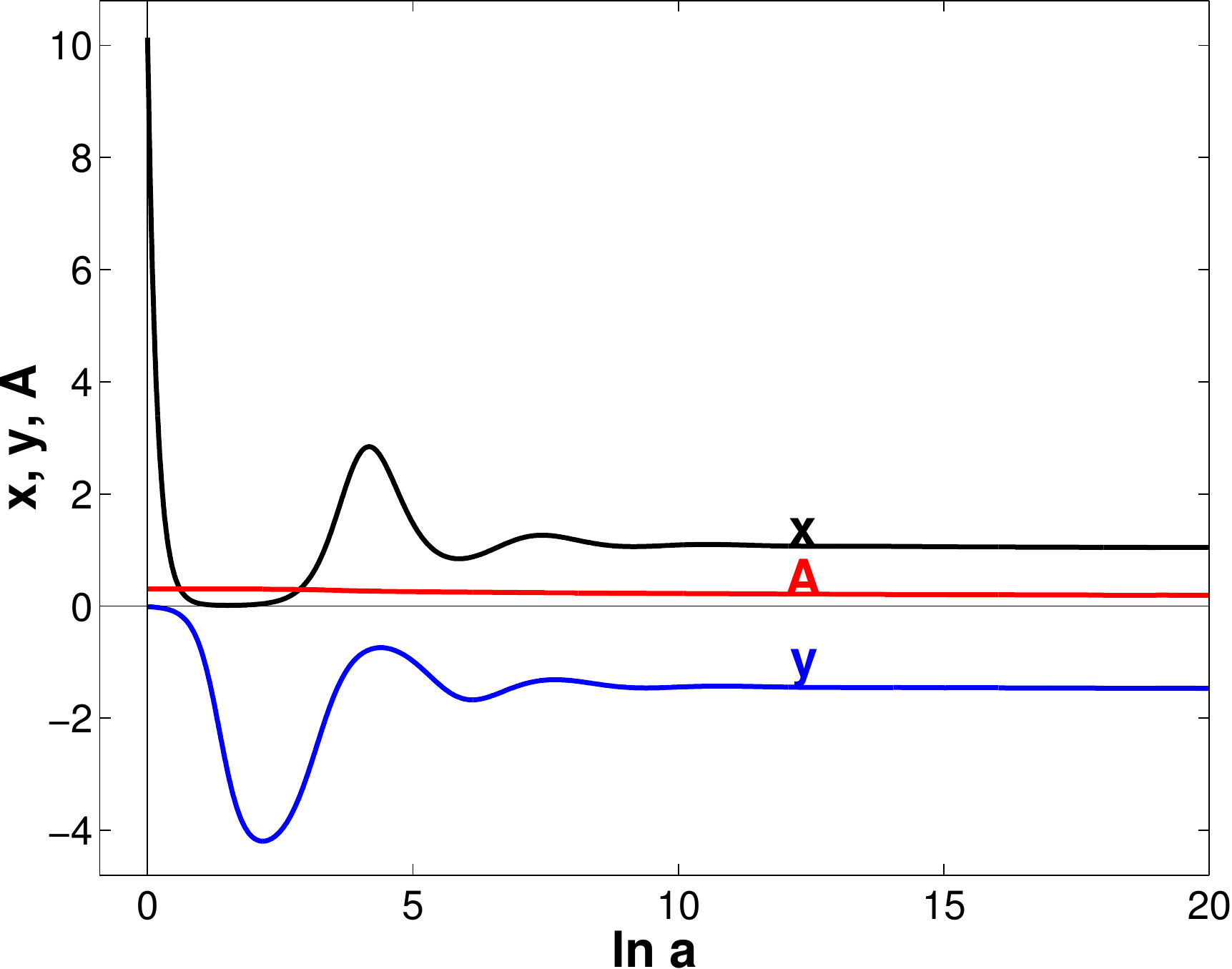}
\caption{(Left panel) Several trajectories in the phase space $(x, y, A)$~~ and ~~(right panel) the evolution of variables $x$,~$y$,~$A$. In the left graph, vertical straight lines of red stars are stationary lines \textbf{1}, \textbf{2}. The separate red stars are the scaling solution point \textbf{3} and the point \textbf{4}. Squares denote initial data, which are chosen as ~~$x(0)=10$,~~ $y(0)=-0.01$,~~ $A(0)$~~ from ~~$0.1$ to $0.3$~~ with step ~~$0.1$~~ for the black curves, ~~$x(0)=10$,~~ $y(0)=-0.01$,~~ $A(0)=0.6$; $0.8$; $1.2$; $1.4$~~ for the blue lines, and ~~$x(0)=10$,~~ $y(0)=-0.01$,~~ $A(0)$~~ from ~~$-0.3$ to $-0.1$~~ with step ~~$0.1$~~ for the green curves. For the right plot, the starting values are $\phi(0)=2.3$,~~ $\dot\phi(0)=1.52$,~~ $H(0)=70$.~~ The parameters are ~~$n=3$, ~~$m=12$, ~~$V_0=1$, ~~$\lambda=1$, ~~$w=0$, ~~${M_{Pl}}^2=1$.}
\label{Fig5}
\end{figure}

\begin{figure}[hbtp] 
\includegraphics[scale=0.46]{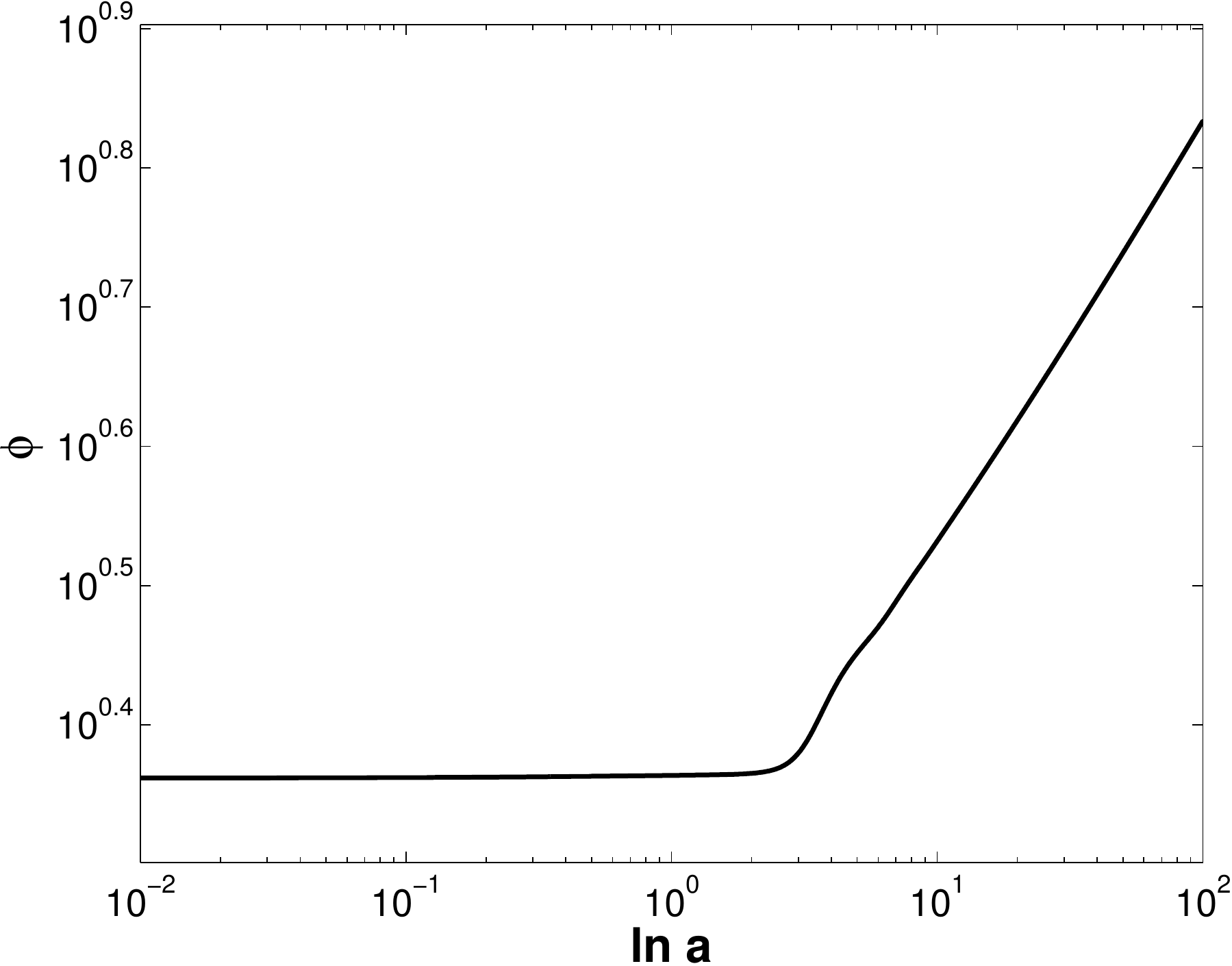}\qquad
\includegraphics[scale=0.46]{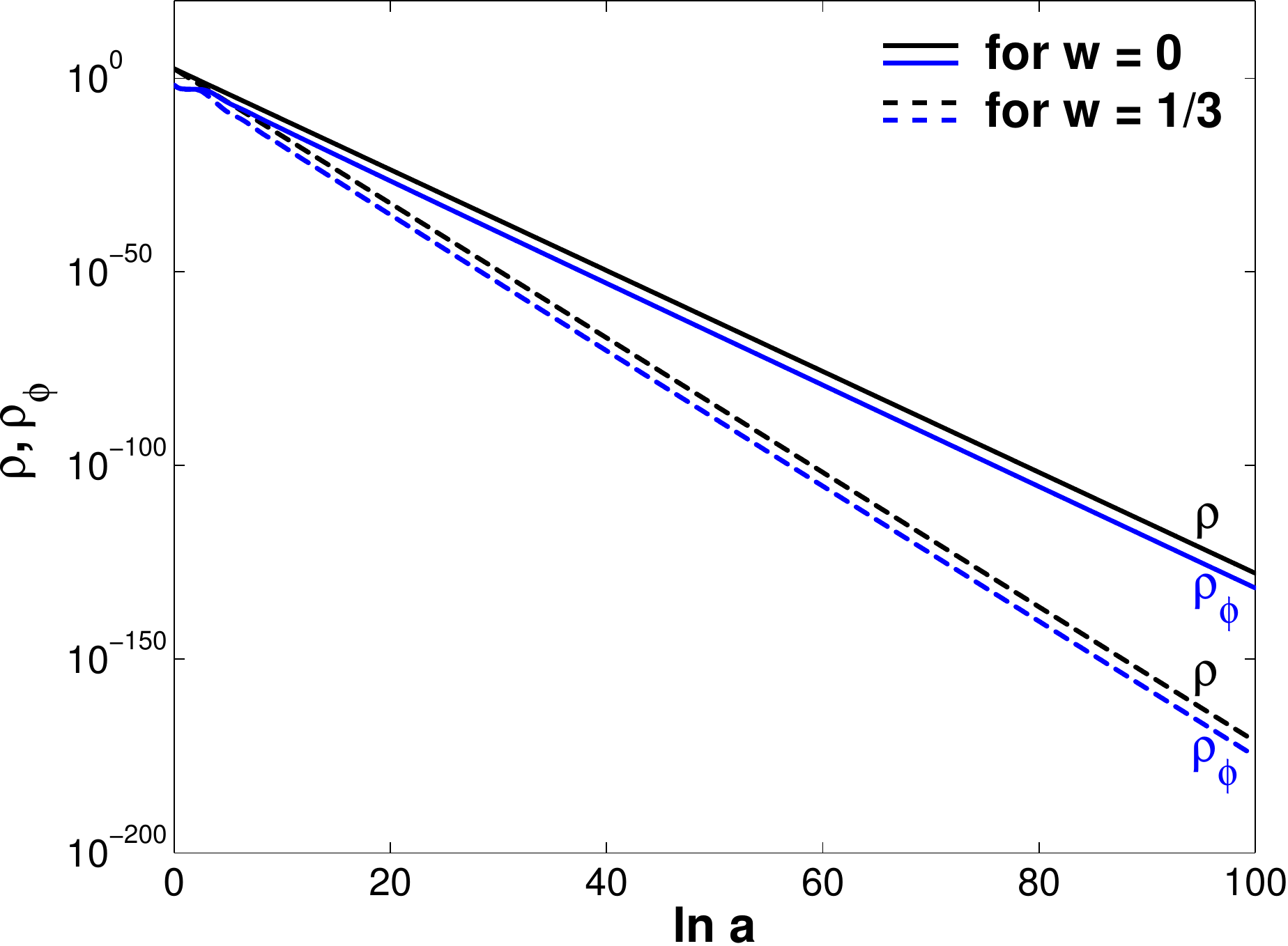}
\caption{The evolution of ~~(left~panel)~$\phi(\ln a)$~~ and ~~(right~panel)~$\rho(\ln a)$~(black), $\rho_{\phi}(\ln a)$~(blue)~~ for the initial data: $\phi(0)=2.3$,~~ $\dot\phi(0)=0.19$,~~ $H(0)=10$.~~ The parameters are ~~$n=3$, ~~$m=7$, ~~$V_0=1$,~~ $\lambda=1$,~~ ${M_{Pl}}^2=1$ ~~for all curves, ~~$w=0$~~ for the solid lines, and ~~$w=1/3$~~ for the dashed lines. For these initial data, the cosmological evolution ends in the asymptotic scaling solution, which corresponds to the fixed point \textbf{3}.}
\label{Fig6}
\end{figure}

\begin{figure}[hbtp] 
\includegraphics[scale=0.32]{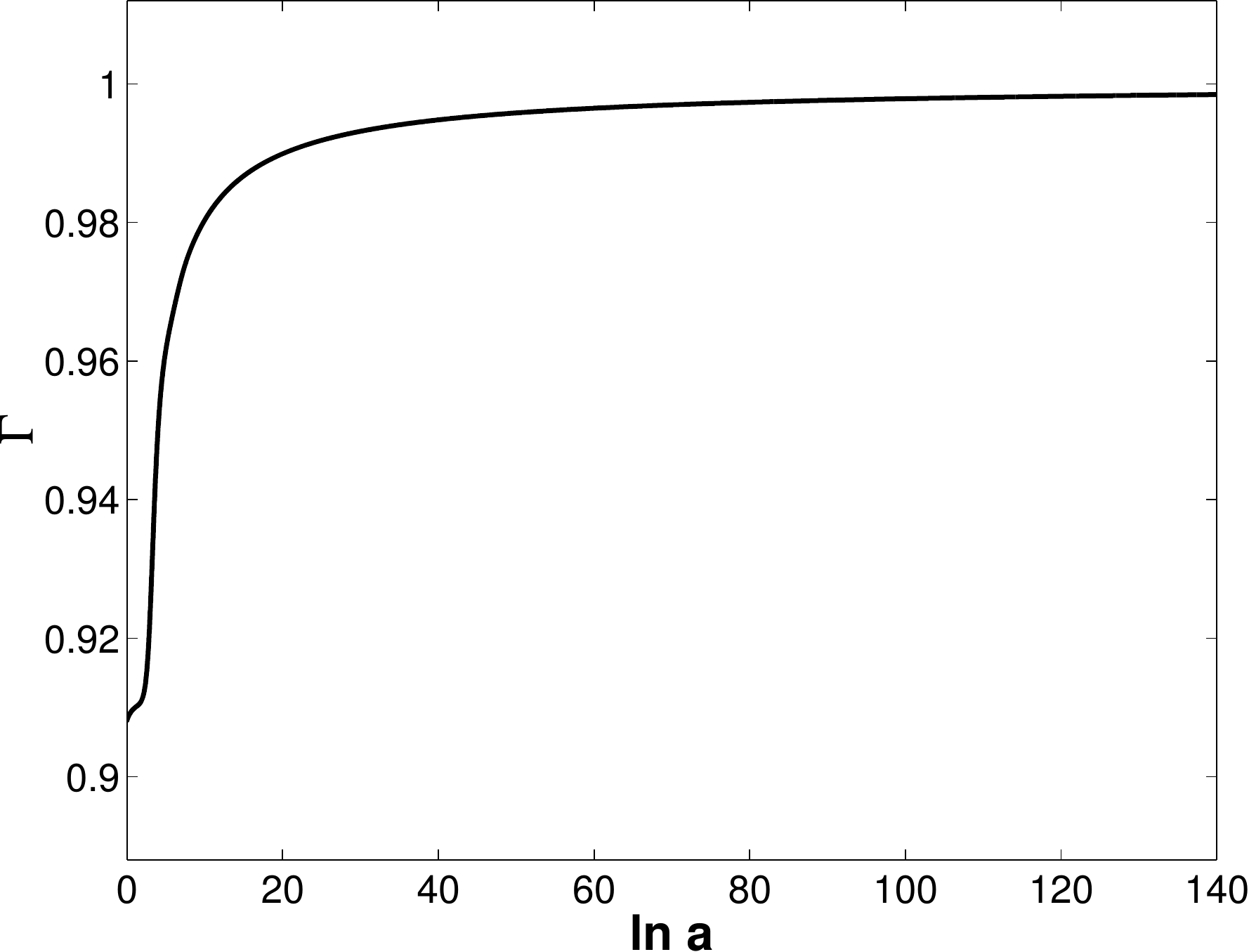}
\includegraphics[scale=0.32]{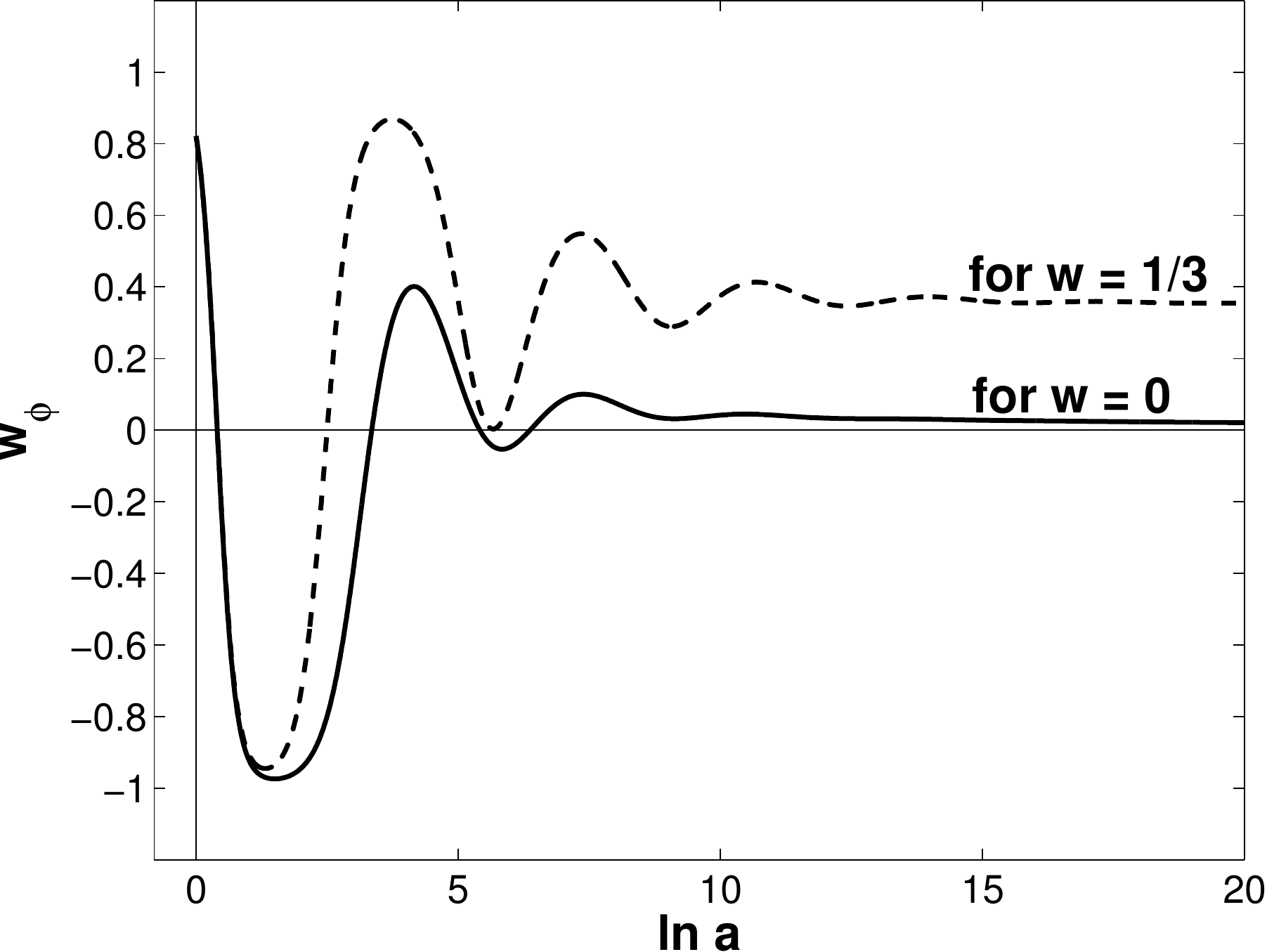}
\includegraphics[scale=0.32]{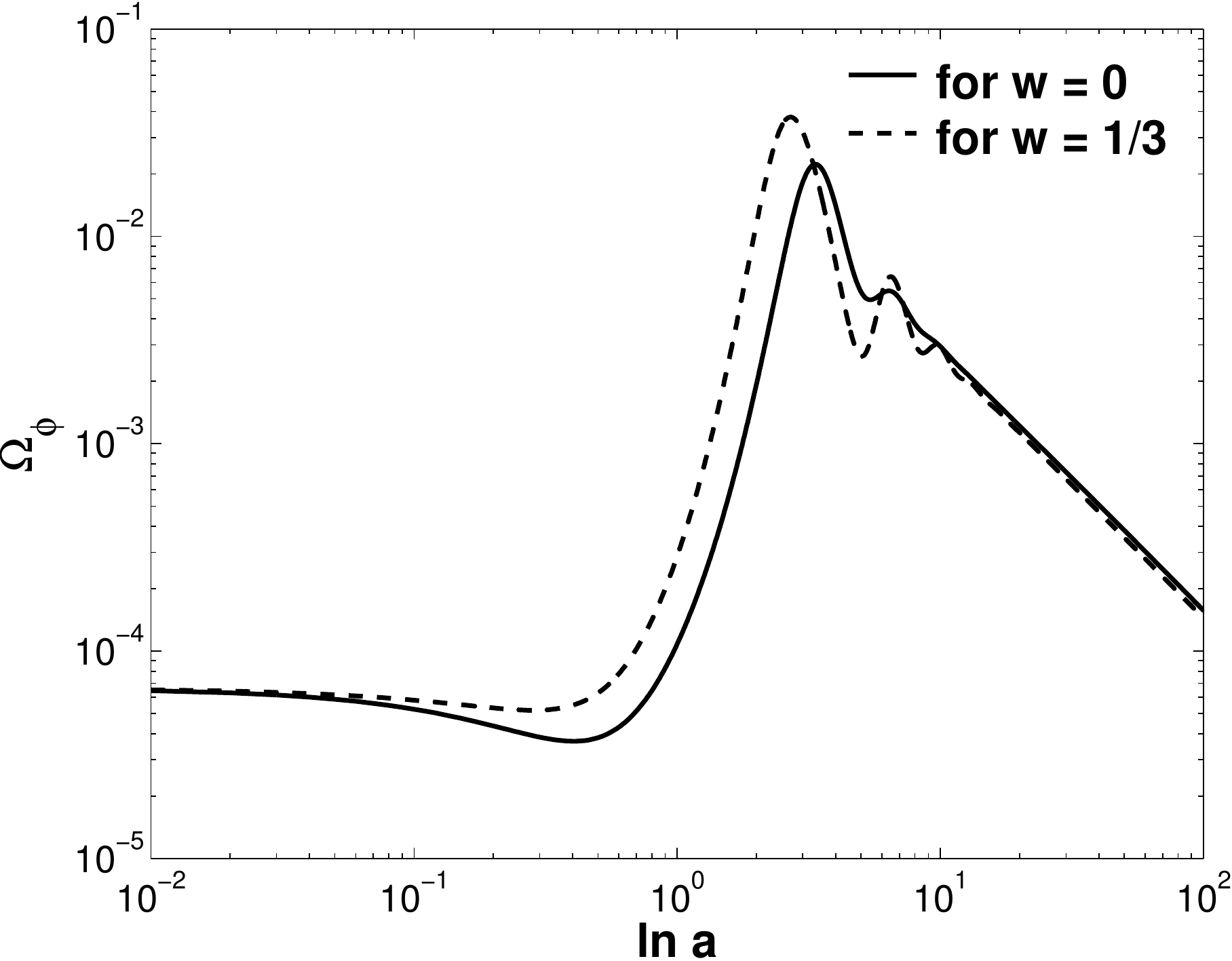}
\caption{The evolution of ~~(left~panel)~$\Gamma(\ln a)$, ~~(middle~panel)~$w_{\phi}(\ln a)$, ~~(right~panel)~$\Omega_{\phi}(\ln a)$~~ for the same initial data and parameters as in Fig.~\ref{Fig6}. For these initial data, the cosmological evolution ends in the asymptotic scaling solution, which corresponds to the fixed point \textbf{3}.}
\label{Fig7}
\end{figure}

\begin{figure}[hbtp] 
\includegraphics[scale=0.32]{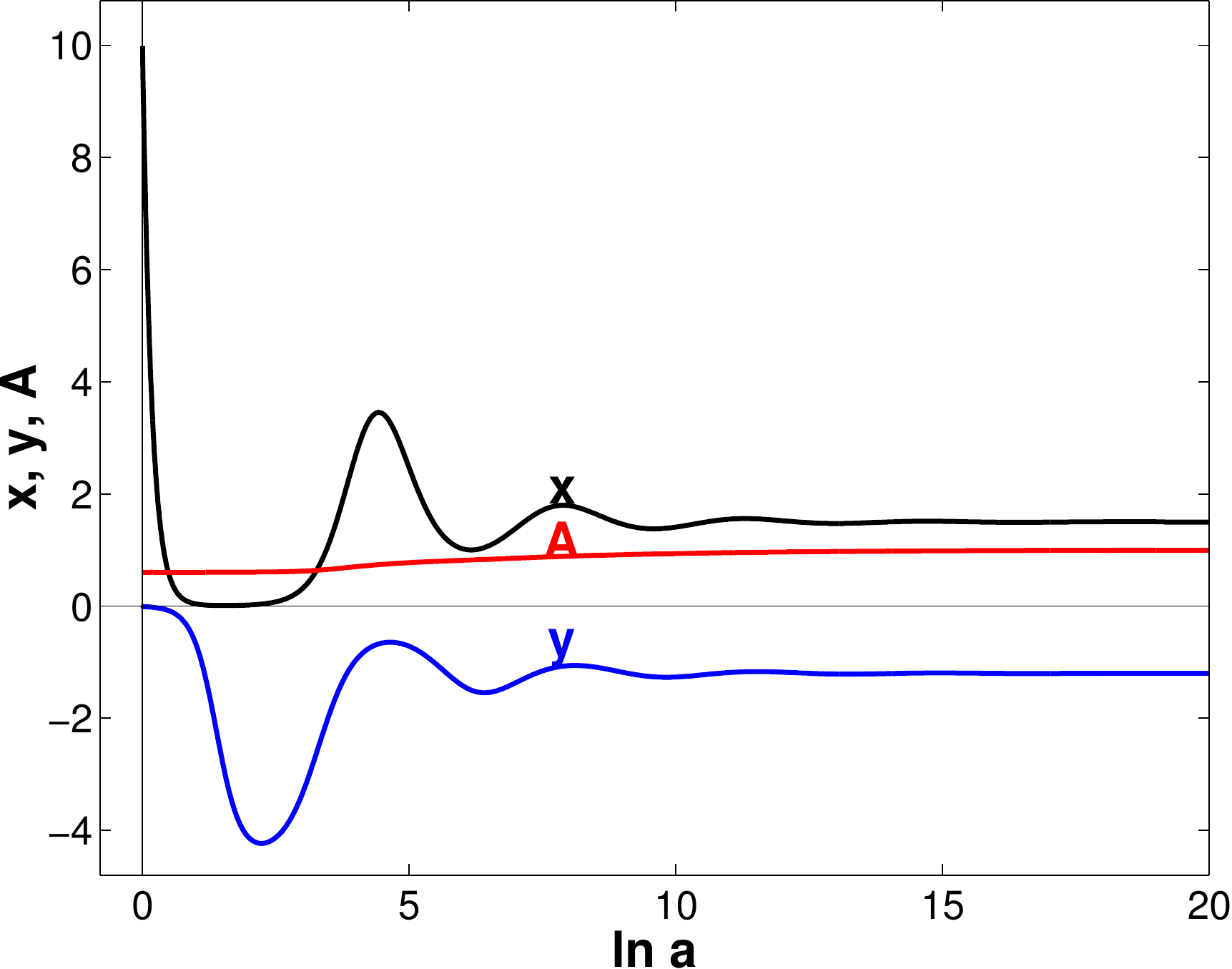}
\includegraphics[scale=0.32]{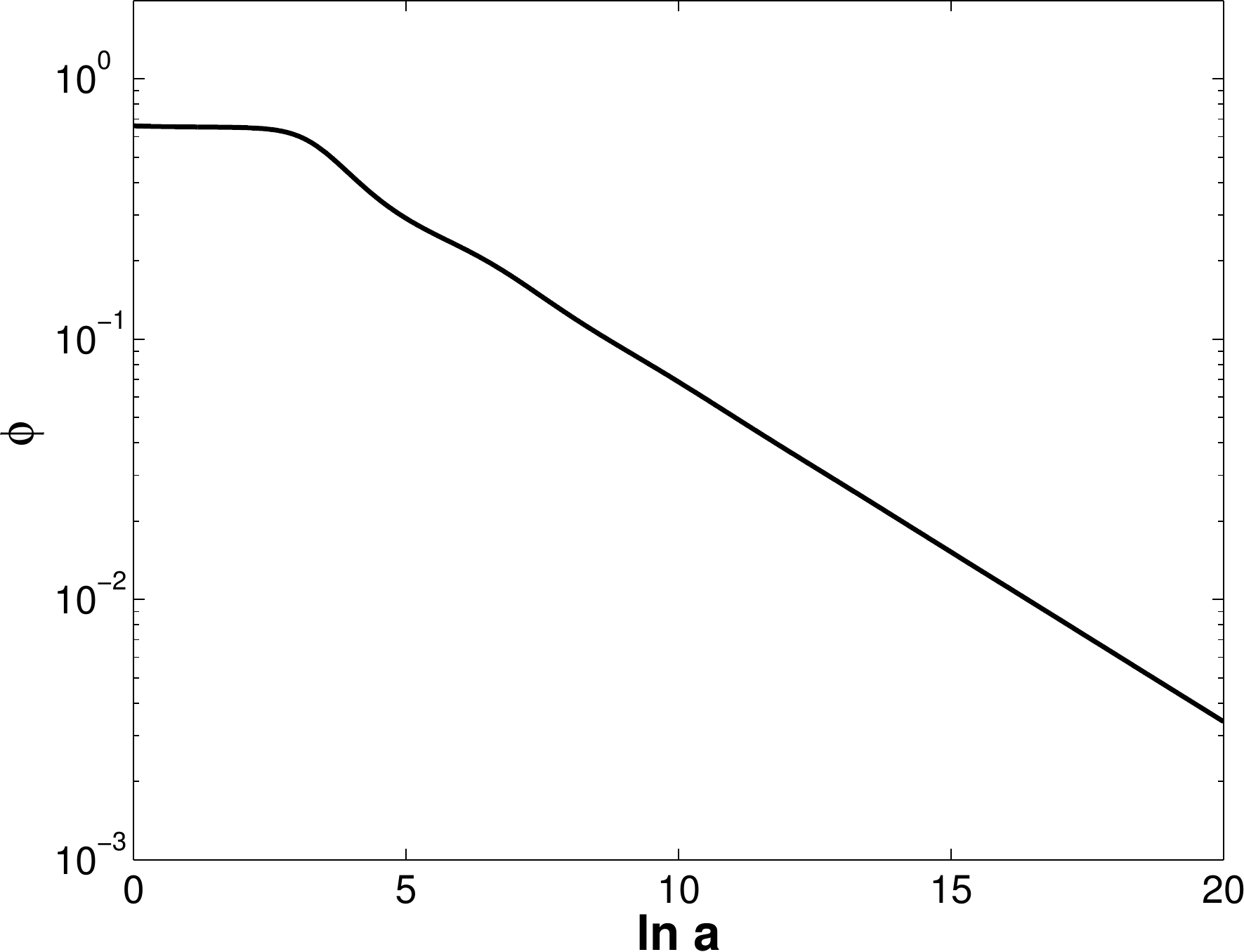}
\includegraphics[scale=0.32]{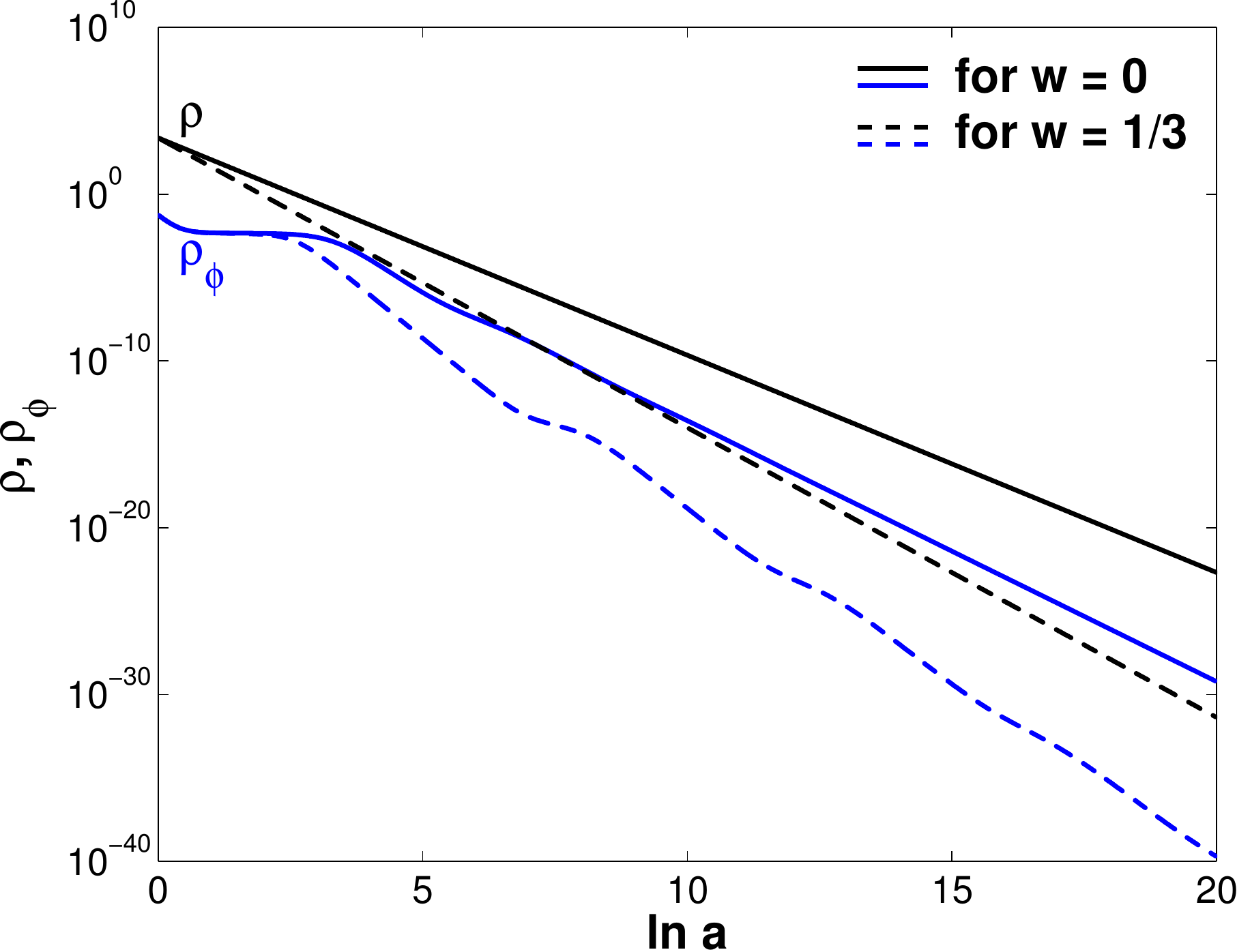}
\caption{The evolution of ~~(left~panel)~variables~$x$,~$y$,~$A$, ~~(middle~panel)~$\phi(\ln a)$, ~~(right~panel)~$\rho(\ln a)$~(black),~$\rho_{\phi}(\ln a)$ (blue)~~ for the initial data: $\phi(0)=0.66$,~~ $\dot\phi(0)=-0.32$,~~ $H(0)=28$.~~ The parameters are ~~$n=3$, ~~$m=12$, ~~$V_0=1$,~~ $\lambda=1$,~~ ${M_{Pl}}^2=1$,~~ $w=0$. For these initial data, the cosmological evolution ends in the asymptotic power-law regime, which corresponds to the fixed point~\textbf{4}.}
\label{Fig8}
\end{figure}

\begin{figure}[hbtp] 
\includegraphics[scale=0.32]{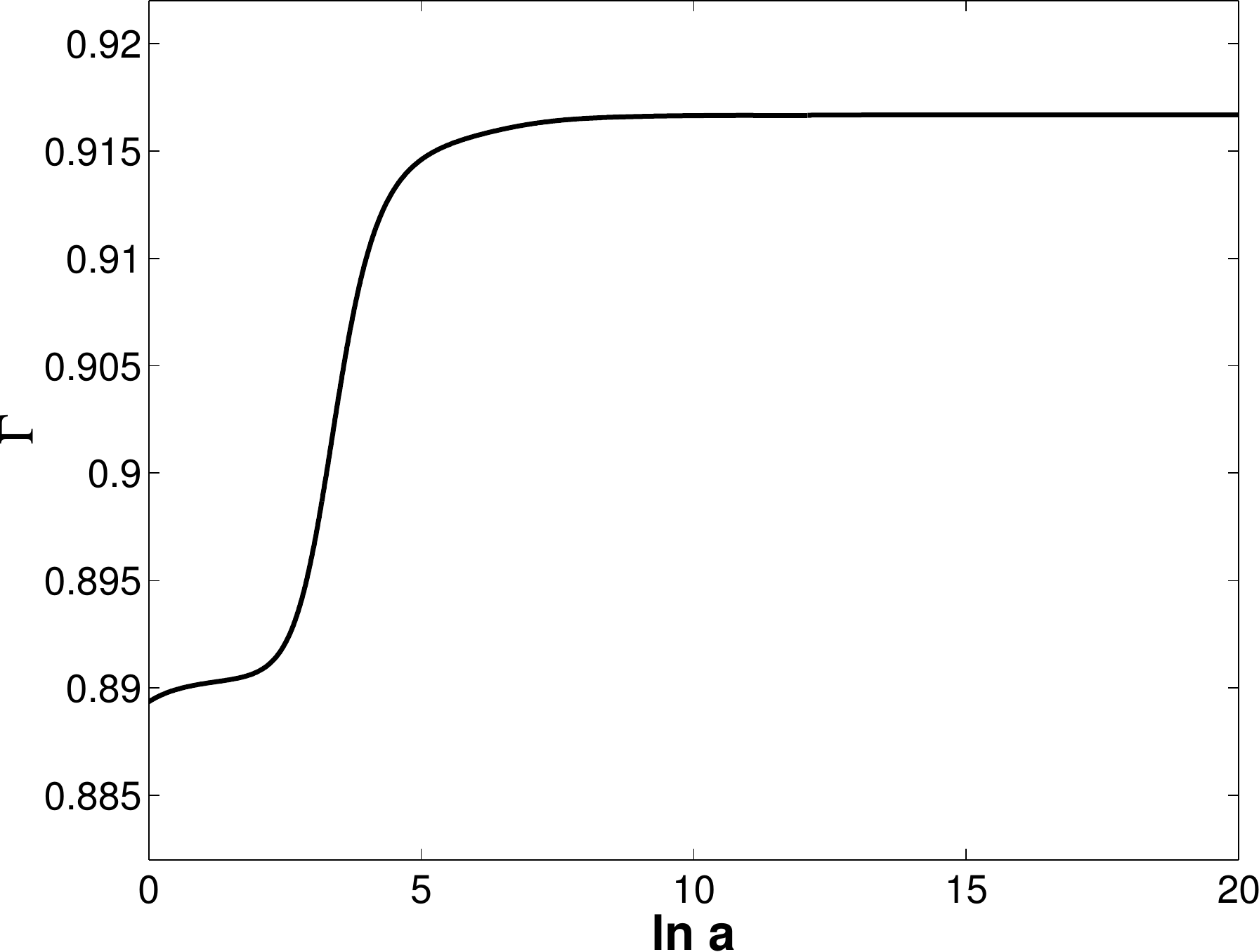}
\includegraphics[scale=0.32]{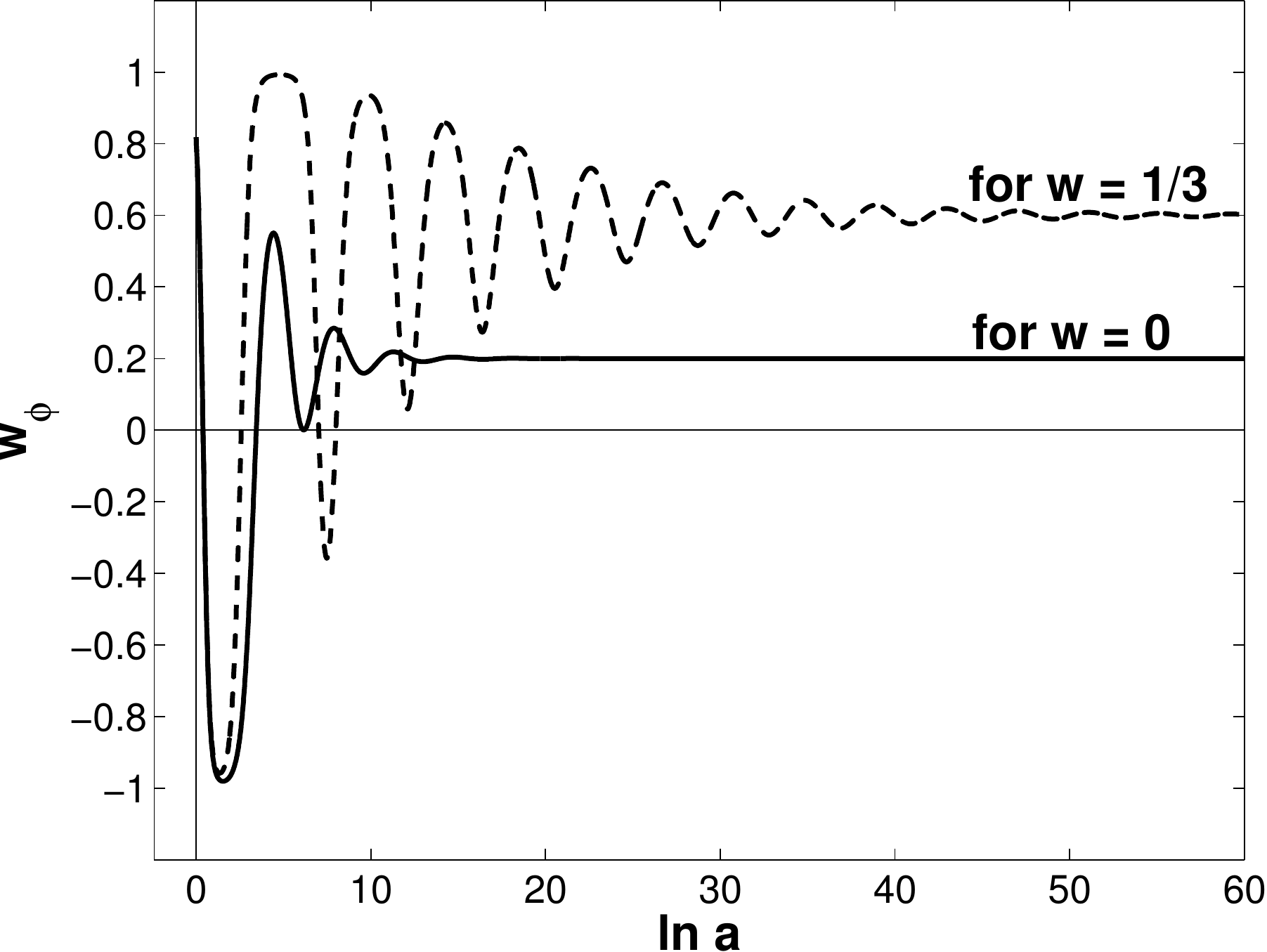}
\includegraphics[scale=0.32]{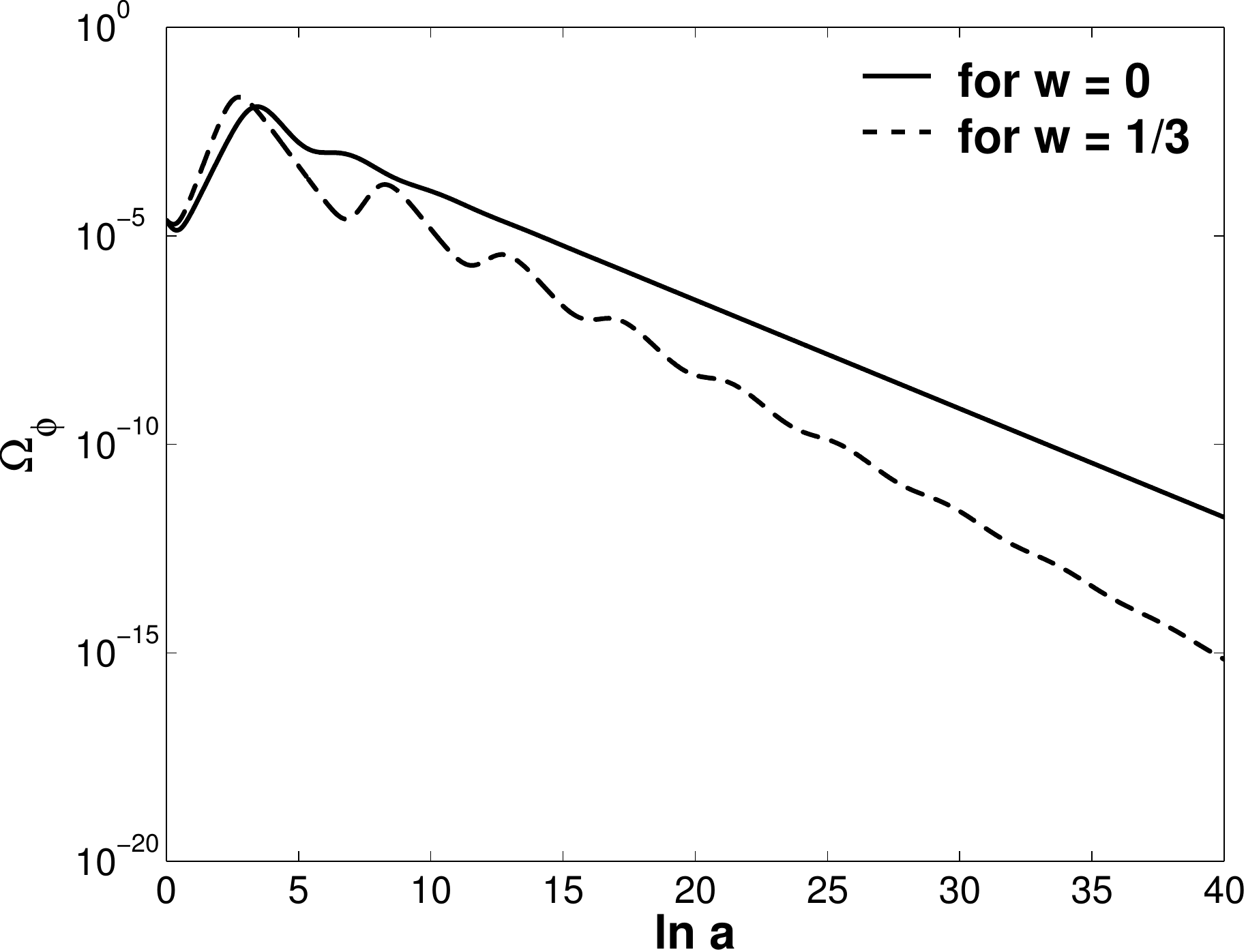}
\caption{The evolution of ~~(left~panel)~$\Gamma(\ln a)$, ~~(middle~panel)~$w_{\phi}(\ln a)$, ~~(right~panel)~$\Omega_{\phi}(\ln a)$~~ for the same initial data and parameters as in Fig.~\ref{Fig8}. For these initial data, the cosmological evolution ends in the asymptotic power-law regime, which corresponds to the fixed point \textbf{4}.}
\label{Fig9}
\end{figure}

    We compare the results obtained from both systems, which are presented in Figs.~\ref{Fig3}-\ref{Fig9}. Let us mention that it is convenient to use the system with original variables also to compute the behavior of certain physical quantities like $\rho_{\phi}$; it also provides a check of results obtained using the autonomous form.
    
    Trajectories in the phase space $(x, y, A)$ are plotted using the system with variables $x$, $y$, $A$ [Eqs.~(\ref{dx})-(\ref{dA})], and they are shown in Figs.~\ref{Fig3}-\ref{Fig5} (left panels). The right panels in Figs.~\ref{Fig3}-\ref{Fig5} and Fig.~\ref{Fig8} (left panel) demonstrate the evolution of variables $x$, $y$, $A$, which are obtained by applying the initial equations of motion (\ref{system11})-(\ref{system31}). In Figs.~\ref{Fig3}-\ref{Fig5} (right panels), we see that the variables $x$, $y$, $A$ approach the coordinates specific to the scaling point for large values of $(\ln a)$ that are at late times. The scaling stationary point~\textbf{3} has the complex type of stability (see the left panel in Fig.~{\ref{Fig5}}). This is stable for some initial data (black trajectories go to point \textbf{3}) and unstable for others (green trajectories leave from the scaling point~\textbf{3} and go to infinity). 
    
    Secondly, for $m>2$, $w\in\left(-1; \frac{m-6}{m+2}\right)$, there exists one more attractor --- namely, the fixed point \textbf{4} in addition to point \textbf{3} (the scaling solution). This case is illustrated in the left panels of Figs.~\ref{Fig4}~and~\ref{Fig5}. Before reaching one of two attractors, the trajectories move from the stationary line \textbf{1} to \textbf{2}, each point of which is a saddle. Then the trajectories in black go to the stable point \textbf{3}, and those in blue go to point \textbf{4}. In Fig.~\ref{Fig8}, the evolution of $x$, $y$, $A$ is shown, which corresponds to the blue curves in Fig.~\ref{Fig5}. The variables $x$,~$y$,~$A$ approach their values in the fixed point \textbf{4} at late times ($\ln a \to +\infty$). 

    Further, integrating the initial system, we plot the dependence of quantities $\phi$, $\rho$, $\rho_{\phi}$, $\Gamma$, $w_{\phi}$, $\Omega_{\phi}$ versus $(\ln a)$ in Figs.~\ref{Fig6}-\ref{Fig9}. When the phase trajectories move near the stationary lines \textbf{1} and \textbf{2}, the scalar field $\phi$ behaves almost like a constant, changing very slowly. It is seen from the left plot in Fig. \ref{Fig6} and the middle graph in Fig. \ref{Fig8} at the initial stages of the evolution $\phi(\ln a)$. We reveal numerically the same power-law dependences of scalar field derivative $\dot\phi$ as in Eqs.~(\ref{phit1}) and (\ref{phit2}), and the behavior of the scalar field $\phi$ near the stationary line \textbf{2}, which coincides with Eq.~(\ref{phi2}), and find that $|\dot\phi|\approx 0$ at both stationary lines. 
 
    For $\ln a\to+\infty$, the scalar field $\phi$, the matter energy density $\rho$ and other said quantities display behaviors corresponding to the asymptotic scaling solution of point \textbf{3} (see Figs.~\ref{Fig6} and \ref{Fig7}) or to the power-law asymptotic regime of point \textbf{4} (see Figs.~\ref{Fig8} and \ref{Fig9}) depending on the initial conditions. We note that a power-law function in double logarithmic scale is a straight line. In the scaling solution, the scalar field $\phi$ increases as ${(\ln a)}^{\alpha}$, the energy densities $\rho$ and $\rho_{\phi}$ decrease with slightly different rates confirming Eqs.~(\ref{rhophi3}) and (\ref{rhophirho3}), the parameter $\Gamma$ tends to unity, the parameter of the scalar field equation of state $w_{\phi}$ approaches $w$, and $\Omega_{\phi}$ decays as ${(\ln a)}^{\beta}$, where $\alpha=\frac{1}{n}$, $\beta=\frac{2(1-n)}{n}$ are power indices calculated from Eq.~(\ref{phi3}) [see formula (\ref{Omegaphi3})].\footnote{Note that, in case of $n=1$ corresponding to the standard exponential potential, $\Omega_{\phi}$ is constant in the scaling regime, as it should be. Secondly, the asymptotic expressions in the scaling regime do not depend upon $m$, as the power law in Eq.~(\ref{pot}) is not relevant in the asymptotic regime, $|\phi|\to\infty$.} In the power-law solution, $\phi$, $\rho$, $\rho_{\phi}$, $\Omega_{\phi}$ decrease as the scale factor in some degrees, and $\Gamma$ and $w_{\phi}$ tend to constants $(m-1)/m$, $(w m+2)/(m-2)$, respectively. Therefore, we conclude that the results of the numerical investigation of the initial system coincide with calculations based upon Eqs.~(\ref{phi})-(\ref{Gamma}), where coordinates of the fixed points \textbf{3} and \textbf{4} are substituted in.  

\section{Conclusion} 
   In this paper, we have investigated cosmological dynamics of scalar field $\phi$ with $V(\phi)\propto {\phi}^m {\rm exp}({-\lambda {\phi}^n/{M_{Pl}}^n})$; $m\geqslant 0$, $n\geqslant1$ in the presence of background matter. We used the autonomous system with three variables $x$, $y$ and $A$, which governs the dynamics of the underlying system. In addition, we also used the dynamical system (\ref{system1})-(\ref{system3}) with original variables; the last was convenient to computing certain physical quantities like $\rho_{\phi}$ and also provided an independent cross-check on results obtained using the autonomous system.
   
   In the case with $m=0$ and $n=1$, for which $\Gamma=V_{\phi \phi}V/V^2_{\phi}=1$, we have a standard scaling solution which tracks the background matter as an attractor. In a general class of potentials with $m\geqslant 0$, $n>1$, we should mimic said behavior in the asymptotic regime as $\Gamma=V_{\phi \phi}V/V^2_{\phi}\to 1$ for $|\phi|\to\infty$. Indeed, in this case, we find three fixed points --- namely, two fixed points given by $x=0$, $y=0$, $A=0$ aand $x=0$, $y=-9/2-3w/2$, $A=0$ which belong to two stationary lines; both of them are unstable solutions. As expected, we also have a scaling solution with $A=0$ corresponding to the third stationary point, which is stable for some limited region of the phase space; see the black trajectories in Figs.~\ref{Fig3}-\ref{Fig5}~(left~panels). The type of stability of the scaling point is complicated. It is demonstrated in Fig.~\ref{Fig5}~(left panel), where the green curves first go to the scaling point, then leave it and go to infinite values of $A$. The scaling solution is reached asymptotically for $|\phi|\to\infty$ such that $w_{\phi}\to w$ ($V(\phi)\to 0$, $\Omega_{\phi}\to 0$). Let us note that $\Omega_{\phi} \propto (\ln a)^{\beta}$, $\beta=2(1-n)/n$ is constant for $n=1$, corresponding to the exponential potential which is a well-known result in the literature. The left plot in Fig. \ref{Fig6} displays the behavior of $\rho_{\phi}$ in the presence of background matter. As seen in the figure, the field being initially subdominant remains frozen till its energy density becomes comparable to the background energy density. Thereafter, the evolution of $\rho_{\phi}$ commences, ultimately catches up with the background, and tracks it forever \textit{\`{a} la} the scaling regime. Our formalism, as demonstrated, is well suited to the investigation of asymptotic scaling solutions. However, the field dominated solution which exists for $m=0$ and $n=1$ is not visible in our analysis though it is there. In the Sec.~\ref{secc}, by specializing  to $n=1$, we explicitly showed the existence and stability of the field dominant solution. This solution might play an important role in the framework of quintessential inflation \cite{Dimopoulos:2017zvq}. 

    The fourth fixed point is also stable for $w\in\left(-1; \frac{m-6}{m+2}\right)$, $m>2$ and corresponds to $A=1$ or $\phi=0$; see the left plots in Figs.~\ref{Fig4} and \ref{Fig5}, which show both fixed points \textbf{3} and \textbf{4}, for which $\Omega_{\phi}\to 0$ and $w_{\rm eff}\to w$. In this case, $w_{\phi} \to (w m+2)/(m-2)$ as the fixed point is reached; for $m=12$, $w=0$, $w_{\phi}\to 0.2$ and for $m=12$, $w=1/3$, $w_{\phi}\to 0.6$ which is clear from Figs.~\ref{Fig8}~and~\ref{Fig9}. It is found that the stability of stationary point \textbf{4} does not depend on the existence of the minimum of the potential at $\phi=0$. For example, when $n=3$,~$m=7$, the scalar field potential does not have the minimum at $\phi=0$, as is shown in Fig.~\ref{Fig2} (left panel), and, however, the stable fixed point \textbf{4} exists for $w<\frac{1}{9}$ (see the left graph in Fig.~\ref{Fig5}). Actually, the minimum of the potential at $\phi=0$ corresponds to the stable scalar field oscillations near zero. In the oscillation regime, the variables $x$, $y$ can go through the infinite values and, therefore, it cannot be revealed from the analysis of the dynamical system with those variables. Let us emphasize that the definitions of autonomous variables are different from the standard ones not suitable for the investigation of the stability of a de Sitter solution; we have added Appendix B, which separately deals with the de Sitter case.\footnote{As demonstrated in Ref.~\cite{Hossain:2014zma}, coupling of the field to massive neutrino matter triggers a minimum in the field potential giving rise to a de Sitter solution which is a late-time attractor of the dynamics.} 
    
    In general, the class of potentials under consideration, is characterized by parameters $n$, $m$ and $\lambda$. In the case of $m=0$, we have the generalized exponential potential [$V\propto {\rm exp}(-\lambda \phi^n/{M^n_{Pl}})$, $n>1$], which has a remarkable property --- namely, it is shallow near the origin and steep for large values of $\phi$. Indeed, the slope of the potential is given by $n\lambda \phi^{n-1}/{M_{Pl}}^{n-1}$; thereby slow roll is ensured (for generic values of $n$ and $\lambda$) provided that $|\phi|\lesssim M_{Pl}$, giving rise to successful inflation for $\lambda\ll 1$ and large values of $n$. In the steep region $|\phi|>M_{Pl}$, scaling behavior is ensured in the asymptotic regime as $\Gamma\to 1$ for $|\phi|\to \infty$. It should further be noted that the asymptotic behavior of $\Gamma$ in the limit $|\phi|\to\infty$ is the same for the generalized exponential potential ($m=0$, $n>1$) and the potential given by expression~(\ref{pot}) ($m\neq 0$, $n>1$); the contribution from the power law in the asymptotic regime is negligible. However, fixed point~\textbf{4} does not exist in the case of the generalized exponential potential. On the other hand, the scaling solution does not exist for $n=0$ (the case of the power-law potential), which is clarified in Appendix A for the framework used by us.\footnote{It should be noted that the singular inflation scenarios exist~\cite{Barrow2} for scalar field models with the power-law potential.} In the general case represented by Eq.~(\ref{pot}), we also have the attractor given by the fixed point {\bf 4}, which, however, is not suitable for late-time acceleration as $\Omega_{\phi}\to 0$, and the evolution around this fixed point is driven by the background matter. In fact, the generalized exponential potential is more useful for said purpose. To this effect, reversing the sign of $\lambda$, we could consider evolution of $\phi$ from large positive values of the field towards the origin. Obviously, in this case one would realize trackerlike behaviour. As for the quintessential inflation, one would require to invoke extra Hubble damping; for instance, such an effect could result from high energy brane corrections capable of supporting slow roll along a steep potential. Unfortunately such a scenario in giving rise to the consistent values of scalar to tensor ratio of perturbations \cite{Sahni:2001qp}. 
   
    During our investigation, we had found that the standard choice of autonomous variables was cumbersome when applied to the potential~(\ref{pot}). As demonstrated in Ref.~\cite{Sami}, the model with $m=0$ can be reconciled easily with observational constraints on inflationary era as well as late-time acceleration, making the latter independent of initial conditions \textit{\`{a} la} the tracking behavior. In the generalized case, the desirable behavior is shown to be intact. We have rigorously shown the existence and stability of a scaling solution using fixed point analysis. Interestingly, we have demonstrated, as expected, that $\Omega_{\phi} \to 0$ as the scaling regime is reached asymptotically; thereby, the nucleosynthesis is taken care of naturally. Indeed, Figs.~\ref{Fig6}~and~\ref{Fig7} clearly show, that $\rho_{\phi}$ tracks the background ultimately, showing the commencement of the scaling regime such that $\Gamma\to 1$, $w_{\phi}\to w$ and $\Omega_{\phi}$ diminishes compared to the background matter content. Said behavior is suited to unified models of inflation and late-time acceleration. 
    
    It should, however, be mentioned that the class of models under consideration cannot on their own account for late-time acceleration. To that effect, we imagine the presence of an additional mechanism responsible for late-time exit from scaling regime to accelerated expansion. As pointed out in Ref.~\cite{Sami}, the interaction with massive neutrino matter may easily trigger said transition at late stages. Finally, we should note that the model based upon the generalized class of potentials has a richer structure than the one considered in Ref.~\cite{Sami} and deserves further attention for model building for quintessential inflation.
    
\section{Acknowledgements}
    We thank T. Padmanabhan for a remark on scaling behavior in the asymptotic regime. M. A. S. is supported by the RFBR grant 18-52-45016. M. S. and N. J. are supported by the Indo-Russia Project (INT/RUS/RFBR/P-315).

\section{Appendix A: the case of $V(\phi)=V_0{\left( \frac{\phi}{M_{Pl}}\right) }^m$ ($n=0$)}
    For $n=0$, the scalar field potential is $V(\phi)=V_0{\left( \frac{\phi}{M_{Pl}}\right) }^m$, and the system (\ref{dx})-(\ref{dA}) has the form
\begin{equation}
\label{dxn0}
\frac{dx}{d(\ln a)}=-2x(3+y+xy),
\end{equation}
\begin{equation}
\label{dyn0}
\frac{dy}{d (\ln a)}=2 x y^2\frac{m-1}{m}+y(9/2+y+3w/2)-\frac{xy^3}{m^2}(x(w-1)+w+1){\left( \frac{1-A}{A}\right) }^2,
\end{equation}
\begin{equation}
\label{dAn0}
\frac{dA}{d (\ln a)}=-\frac{2}{m}xyA(1-A).
\end{equation}
    
We solve this system and find the following stationary points.
\\
\\\textbf{1.} $x=0$, $y=0$, $A\in(-\infty; +\infty)$.
\\This is stationary line. Eigenvalues are calculated as
 \begin{equation}
\begin{array}{l}
L_1=-6<0,\\
L_2=9/2+3w/2>0 \text{~~~~for $w\in[-1; 1]$,}\\
L_3=0.
\end{array}
\end{equation}
\\
\\
\\\textbf{2.} $x=0$, $y=-\frac{3}{2}(w+3)$, $A\in(-\infty; +\infty)$.
\\We find eigenvalues for this stationary line:
 \begin{equation}
\begin{array}{l}
L_1=3+3w\geqslant0 \text{~~~~for $w\in[-1; 1]$,}\\
L_2=-9/2-3w/2<0 \text{~~~~for $w\in[-1; 1]$,}\\
L_3=0.
\end{array}
\end{equation}
\\
\\
\\\textbf{3.} $x=-\frac{m(1+w)}{wm-m+4}$, $y=\frac{3(wm-m+4)}{2(m-2)}$, $A=1$.
\\This stationary point exists for $m\neq0$, $m\neq2$, $w\neq\frac{m-4}{m}$. Eigenvalues are found as
\begin{equation}
\begin{array}{l}
L_1=-\frac{3(1+w)}{m-2}<0 \text{~~~~for $m>2$, $w\in(-1;1]$,}\\
L_{2,3}=\frac{3}{4(m-2)}\left( f_1(m,w)\pm\sqrt{f_2(m,w)}\right) \text{~~~~$Re(L_{2,3})<0$~~ for $m>2$, $w\in\left(-1; \frac{m-6}{m+2}\right) $,}
\end{array}
\end{equation}
where $f_1(m,w)=w(m+2)-m+6$,
\\\text{~~~~~~~~}$f_2(m,w)=(9m^2-12m+4)w^2+2(20m-m^2-20)w-7m^2+36m-28$.

    We see that the coordinates $x$ and $y$ of these stationary points are not equal to $\frac{1+w}{1-w}$ and $\frac{3}{2}(w-1)$, respectively. Therefore, the scaling solution does not exist in the model with the potential $V(\phi)=V_0{\left( \frac{\phi}{M_{Pl}}\right) }^m$. 

\section{Appendix B: the stability of the $\text{\bf de}$ Sitter solution}    
    The de Sitter solution $H=H_0$, $\phi=\phi_0$, $\rho=0$ exists for 
\begin{equation}
\begin{array}{c}
\label{deSitter}  
V_{\phi}(\phi_0)=V_0 \frac{{\phi_0}^{m-1}}{{M_{Pl}}^m}e^{-\lambda\frac{{\phi_0}^n}{{M_{Pl}}^n}}\left( m-\lambda n\frac{{\phi_0}^n}{{M_{Pl}}^n}\right)=0 ~~\Rightarrow~~ \textbf{(1)~} \phi_0=0,~~ \textbf{(2)~} {\phi_0}^n=\frac{m{M_{Pl}}^n}{\lambda n},\\ 
3{H_0}^2{M_{Pl}}^2=V_0 {\left( \frac{\phi_0}{M_{Pl}}\right) }^m e^{-\lambda\frac{{\phi_0}^n} {{M_{Pl}}^n}} ~~\Rightarrow ~~\textbf{(1)~} H_0=0,~~ \textbf{(2)~} {H_0}^2=\frac{V_0}{3{M_{Pl}}^2} {\left( \frac{m}{\lambda n}\right)}^{\frac{m}{n}}e^{-\frac{m}{n}}.
\end{array}
\end{equation}
Adding small perturbations to the solution $\phi(t)=\phi_0+\delta\phi$, $\dot\phi(t)=\delta\dot\phi$, $H(t)=H_0+\delta H$, we substitute these into Eqs.~(\ref{system2}) and (\ref{system3}) and find, in the linear approach,
\begin{equation}
\label{system22}
\delta\dot H=-3{H_0}(1+w)\delta H +\frac{V_{\phi}(\phi_0)(1+w)}{2{M_{Pl}}^2}\delta\phi,
\end{equation}
\begin{equation}
\label{system32}
\delta\ddot\phi+3H_0\delta\dot\phi+V_{\phi\phi}(\phi_0)\delta\phi=0,
\end{equation}    
where $\rho=3H^2{M_{Pl}}^2-\frac{1}{2}{\dot\phi}^2-V(\phi)$ has been substituted in. We introduce new variables $s_1=\delta\phi$, $s_2=\delta\dot\phi$, $s_3=\delta H$ and obtain a first-order system of differential equations:
\begin{equation}
\label{ds1}
\dot s_1=s_2,
\end{equation}
\begin{equation}
\label{ds2}
\dot s_2=-3H_0 s_2-V_{\phi\phi}(\phi_0)s_1,
\end{equation}
\begin{equation}
\label{ds3}
\dot s_3=-2H_0(1+w)s_3+\frac{V_{\phi}(\phi_0)(1+w)}{2{M_{Pl}}^2}s_1.
\end{equation}
The eigenvalues for the matrix of this system are
\begin{equation}
\begin{array}{l}
L_1=-3H_0(1+w)<0 \text{~~~~for $H_0>0$, $w\in(-1; 1]$,}
\\L_{2,3}=\frac{3}{2}\left( -H_0\pm\sqrt{{H_0}^2-4V_{\phi\phi}(\phi_0)/9}\right)   \text{~~~~$Re(L_{2,3})<0$~~ for $H_0>0$, $V_{\phi\phi}(\phi_0)>0$.}
\end{array}
\end{equation}
Therefore, we have a stable de Sitter solution in the minimum of the potential at $\phi=\phi_0$ (that is, for $V_{\phi\phi}(\phi_0)>0$).

\section{Appendix C: Fixed point analysis for the case of $m=0$ and $n=1$}
\label{appendC}
    To find the fixed points or critical points of the dynamical system (\ref{dx2}), (\ref{dy2}), we set their left-hand sides to zero,
\begin{eqnarray}
\label{dx3}
\frac{dx}{d(\ln a)}=f(x,y)\vert_{(x_c, y_c)}=0, \\
\label{dy3}
\frac{dy}{d(\ln a)}=g(x,y)\vert_{(x_c, y_c)}=0,
\end{eqnarray}
which allows us to obtain a set of critical points ($x_c, y_c$),
\begin{eqnarray}
1a &:& x=\frac{\lambda ^2}{6-\lambda ^2},~~ y=\frac{1}{2} \left(\lambda ^2-6\right)\\
1b&:& x=0,~~ y=0\\
2a&:& x=\frac{1+w}{1-w},~~ y=\frac{3}{2}(w-1)\\
2b &:& x=0,~~ y=-\frac{3}{2}(w+3)
\end{eqnarray}
To check for the stability of the critical points, we do the linear perturbation \cite{nonlinear,Copeland:2006wr}, $\delta x$ and $\delta y$, around the critical points $x_c$ and $y_c$,
\begin{eqnarray}
\label{linpert1}
x &=& x_c+\delta x,~~~~ y= y_c+\delta y.
\end{eqnarray}
Substituting Eq.~(\ref{linpert1}) back into Eqs.~(\ref{dx3}) and (\ref{dy3}), we get the following set of perturbation equations,
\begin{equation}
\frac{d}{d\ln a}\binom{\delta x}{\delta y}= \mathcal{M} \binom{\delta x}{\delta y},
\end{equation}
where
\begin{equation}
\mathcal{M}=\begin{pmatrix} 
\frac{\partial f}{\partial x} & \frac{\partial f}{\partial y}\\ 
\frac{\partial g}{\partial x} & \frac{\partial g}{\partial y} 
\end{pmatrix}\Bigg{\vert}_{x_c, y_c}
\end{equation}
is the perturbation matrix. The nature of the eigenvalues of the matrix $\mathcal{M}$ gives the stability criteria of the system; see Refs.~\cite{nonlinear,Copeland:2006wr} for details. 
In this appendix, we shall consider points \textbf{1a} and \textbf{2a} only, as they are central to our discussion. 
\\
\\\textbf{(1a)}
In this case, the eigenvalues are
\begin{equation}
\begin{array}{l}
L_1=\frac{1}{2}(\lambda^2-6),
\\L_2=\lambda^2-3(1+w).
\end{array}
\end{equation}
It is clear that ~~$L_1<0$, $L_2<0$~~ for $\lambda^2<3(1+w)$ \,\ ($0\leqslant w<1$). This is the field dominant solution with $\Omega_{\phi}=1$ which was invisible in our general analysis. This solution is a global attractor and does not exist for $m=0$ and $n>1$.
\\
\\\textbf{(2a)} For this point, the eigenvalues are 
\begin{equation}
\label{L12}
L_{1,2}=\frac{3}{4}(w-1)\pm\frac{3}{4\lambda} \sqrt{(w-1)\left( \lambda^2 (9 w+7)-24 {(w+1)}^2\right) }.
\end{equation}
Since $\Omega_{\phi}<1$, we have $\lambda^2>3(1+w)$. Taking into account the fact that $24(1+w)^2/(9w+7)>3(w+1)$ for $0\leqslant w<1$\footnote{Here we consider the realistic background matter.} we find that this fixed point is an attractor.\footnote{Interestingly, Eqs.~(\ref{dx2}) and (\ref{dy2}) look different in the details than the autonomous equations in Ref.~\cite{Copeland} but lead to the same conclusion.}  Namely, we have the following:
\\
\\(1) For ~~${\lambda}^2>24(1+w)^2/(9w+7)$~~ it is a stable focus, as the expression under the square root in $L_1$ and $L_2$ [see Eq.~(\ref{L12})] is negative, and the real parts of both the eigenvalues are equal and negative.
\\
\\(2) For ~~$3(1+w)<\lambda^2\leqslant 24(1+w)^2/(9w+7)$,~~ it is a stable node, as the eigenvalues $L_1$, $L_2$ do not have an imaginary part, and both of them are negative.  


\begin{thebibliography}{99}
\bibitem{Copeland} 
  E.~J.~Copeland, A.~R.~Liddle, and D.~Wands,
  Exponential potentials and cosmological scaling solutions,
  Phys.\ Rev.\ D {\bf 57},
  4686 (1998).

\bibitem{Ferreira:1997au} 
  P.~G.~Ferreira and M.~Joyce,
  Structure formation with a self-tuning scalar field,
  Phys.\ Rev.\ Lett. {\bf 79},
  4740 (1997). 

\bibitem{Sami:2004ic} 
  M.~Sami and N.~Dadhich,
  Unifying brane world inflation with quintessence,
  TSPU Bull. {\bf 7}, 25 (2004).
  
\bibitem{Sahni:2001qp} 
  V.~Sahni, M.~Sami, and T.~Souradeep,
  Relic gravity waves from brane world inflation,
  Phys.\ Rev.\ D {\bf 65}, 
  023518 (2001).

\bibitem{Tashiro:2003qp} 
  H.~Tashiro, T.~Chiba, and M.~Sasaki,
  Reheating after quintessential inflation and gravitational waves,
  Classical Quantum Gravity {\bf 21}, 
  1761 (2004).
  
\bibitem{Hossain:2014zma} 
  M.~W.~Hossain, R.~Myrzakulov, M.~Sami, and E.~N.~Saridakis,
  Unification of inflation and dark energy à la quintessential inflation,
  Int.\ J.\ Mod.\ Phys.\ D {\bf 24}, 
  1530014 (2015).
  
\bibitem{Geng:2015fla} 
  C.~Q.~Geng, M.~W.~Hossain, R.~Myrzakulov, M.~Sami, and E.~N.~Saridakis,
  Quintessential inflation with canonical and noncanonical scalar fields and Planck 2015 results,
  Phys.\ Rev.\ D {\bf 92}, 
  023522 (2015).

\bibitem{Linder:2008pp} 
  E.~V.~Linder,
  Mapping the cosmological expansion,
  Rept.\ Prog.\ Phys.\ {\bf 71}, 056901 (2008).

\bibitem{thawing1}
  M.~Sami, 
  A primer on problems and prospects of dark energy, 
  Curr.\ Sci.\ {\bf 97}, 887 (2009).

\bibitem{Sami} 
  C.-~Q.~Geng, C.-~C.~Lee, M.~Sami, E.~N.~Saridakis, and A.~A.~Starobinsky,
  Observational constraints on successful model of quintessential inflation,
  J.\ Cosmol.\ Astropart.\ Phys.\ {\bf 06} (2017) 011.

\bibitem{Peebles:1998qn} 
  P.~J.~E.~Peebles and A.~Vilenkin,
  Quintessential inflation,
  Phys.\ Rev.\ D {\bf 59}, 
  063505 (1999).

\bibitem{Spokoiny:1993kt} 
  B.~Spokoiny,
  Deflationary Universe scenario,
  Phys.\ Lett.\ B {\bf 315}, 40 (1993).

\bibitem{Peebles:1999fz} 
  P.~J.~E.~Peebles and A.~Vilenkin,
  Noninteracting dark matter,
  Phys.\ Rev.\ D {\bf 60}, 
  103506 (1999).

\bibitem{Peloso:1999dm} 
  M.~Peloso and F.~Rosati,
  On the construction of quintessential inflation models,
  J.\ High Energy Phys.\ {\bf 12} (1999) 026.

\bibitem{Dimopoulos:2000md} 
  K.~Dimopoulos,
  Towards a model of quintessential inflation,
  Nucl.\ Phys.\ Proc.\ Suppl.\ {\bf 95}, 70 (2001).

\bibitem{Copeland:2000hn} 
  E.~J.~Copeland, A.~R.~Liddle, and J.~E.~Lidsey,
  Steep inflation: Ending brane world inflation by gravitational particle production,
  Phys.\ Rev.\ D {\bf 64}, 
  023509 (2001).
  
\bibitem{Majumdar:2001mm} 
  A.~S.~Majumdar,
  From brane assisted inflation to quintessence through a single scalar field,
  Phys.\ Rev.\ D {\bf 64}, 
  083503 (2001).

\bibitem{Sami:2004xk} 
  M.~Sami and V.~Sahni,
  Quintessential inflation on the brane and the relic gravity wave background,
  Phys.\ Rev.\ D {\bf 70}, 
  083513 (2004).

\bibitem{Rosenfeld:2005mt} 
  R.~Rosenfeld and J.~A.~Frieman,
  A simple model for quintessential inflation,
  J.\ Cosmol.\ Astropart.\ Phys.\ {\bf 09} (2005) 003.
 
\bibitem{Dimopoulos:2001ix} 
  K.~Dimopoulos and J.~W.~F.~Valle,
  Modeling quintessential inflation,
  Astropart.\ Phys.\ {\bf 18} (2002) 287.

\bibitem{Giovannini:2003jw} 
  M.~Giovannini,
  Low-scale quintessential inflation,
  Phys.\ Rev.\ D {\bf 67}, 
  123512 (2003).
  
\bibitem{Tsujikawa:2013fta} 
  S.~Tsujikawa,
  Quintessence: A Review,
  Classical Quantum Gravity {\bf 30}, 
  214003 (2013).

\bibitem{Hossain:2014xha} 
  M.~W.~Hossain, R.~Myrzakulov, M.~Sami, and E.~N.~Saridakis,
  Variable gravity: A suitable framework for quintessential inflation,
  Phys.\ Rev.\ D {\bf 90}, 
  023512 (2014).
  
\bibitem{Dimopoulos:2017zvq} 
  K.~Dimopoulos and C.~Owen,
  Quintessential inflation with $\alpha$-attractors,
  J.\ Cosmol.\ Astropart.\ Phys.\ {\bf 06} (2017) 027.
  
\bibitem{Ahmad:2017itq} 
  S.~Ahmad, R.~Myrzakulov, and M.~Sami,
  Relic gravitational waves from quintessential inflation,
  Phys.\ Rev.\ D {\bf 96}, 
  063515 (2017).

\bibitem{Linde} 
  Y.~Akrami, R.~Kallosh, A.~Linde, and V.~Vardanyan,
  Dark energy, $\alpha$-attractors, and large-scale structure surveys,
  J.\ Cosmol.\ Astropart.\ Phys.\ {\bf 06} (2018) 041.

\bibitem{Dimopoulos:2018} 
  K.~Dimopoulos, L.~D.~Wood, and C.~Owen,
  Instant preheating in quintessential inflation with $\alpha$-attractors,
  Phys.\ Rev.\ D {\bf 97}, 
  063525 (2018).
  
\bibitem{Jaman:2018ucm} 
  N.~Jaman and K.~Myrzakulov,
  Braneworld inflation with effective $\alpha$-attractor potential,
  Phys.\ Rev.\ D {\bf 99}, 
  103523 (2019).

\bibitem{Hossain:2018pnf} 
  M.~W.~Hossain,
  Quintessential inflation: A unified scenario of inflation and dark energy,
  EPJ Web Conf.\ {\bf 168}, 04007 (2018).
  
\bibitem{Cardenas:2006py} 
  V.~H.~Cardenas,
  Tachyonic quintessential inflation,
  Phys.\ Rev.\ D {\bf 73}, 
  103512 (2006).
  
\bibitem{Sanchez1} 
  J.~C.~B. Sanchez and K.~Dimopoulos,
  Trapped quintessential inflation,
  Phys.\ Lett.\ B {\bf 642}, 
  294 (2006);
  {\bf 647}, 
  526 (E) (2007).

\bibitem{Sanchez2}  
  J.~C.~B. Sanchez and K.~Dimopoulos,
  Trapped quintessential inflation in the context of flux compactifications,
  J.\ Cosmol.\ Astropart.\ Phys.\ {\bf 10} (2007) 002.

\bibitem{Neupane:2007mu} 
  I.~P.~Neupane,
  Reconstructing a model of quintessential inflation,
  Classical Quantum Gravity {\bf 25}, 
  125013 (2008).

\bibitem{BasteroGil:2009eb} 
  M.~Bastero-Gil, A.~Berera, B.~M.~Jackson, and A.~Taylor,
  Hybrid quintessential inflation,
  Phys.\ Lett.\ B {\bf 678}, 
  157 (2009).

\bibitem{Guendelman:2016kwj} 
  E.~Guendelman, E.~Nissimov, and S.~Pacheva,
  Quintessential inflation, unified dark energy and dark matter, and Higgs mechanism,
  Bulgarian Journal of Physics {\bf 44}, 
  015 (2017).

\bibitem{Haro1} 
  J.~Haro,
  Different reheating mechanisms in quintessence inflation,
  Phys.\ Rev.\ D {\bf 99}, 
  043510 (2019). 

\bibitem{Haro2}  
  J.~Haro, W.~Yang, and S.~Pan, 
  Reheating in quintessential inflation via gravitational production of heavy massive particles: A detailed analysis,
  J.\ Cosmol.\ Astropart.\ Phys.\ {\bf 01} (2019) 023.

\bibitem{Steinhardt:1999nw} 
  P.~J.~Steinhardt, L.~Wang, and I.~Zlatev,
  Cosmological tracking solutions,
  Phys.\ Rev.\ D {\bf 59}, 
  123504 (1999).

\bibitem{Ferreira:1997hj} 
  P.~G.~Ferreira and M.~Joyce,
  Cosmology with a primordial scaling field,
  Phys.\ Rev.\ D {\bf 58}, 
  023503 (1998).
  
\bibitem{Ratra:1987rm} 
  B.~Ratra and P.~J.~E.~Peebles,
  Cosmological consequences of a rolling homogeneous scalar field,
  Phys.\ Rev.\ D {\bf 37}, 
  3406 (1988).

\bibitem{Guo:2003eu} 
  Z.-~K.~Guo, Y.-~S.~Piao, and Y.-~Z.~Zhang,
  Cosmological scaling solutions and multiple exponential potentials,
  Phys.\ Lett.\ B {\bf 568}, 1 (2003).

\bibitem{Copeland:2006wr} 
  E.~J.~Copeland, M.~Sami, and S.~Tsujikawa,
  Dynamics of dark energy,
  Int.\ J.\ Mod.\ Phys.\ D {\bf 15}, 
  1753 (2006). 

\bibitem{Singh}   
   P.~Singh, M.~Sami, and N.~Dadhich, 
   Cosmological dynamics of a phantom field,
   Phys.\ Rev.\ D {\bf 68}, 
   023522 (2003).
  
\bibitem{Wetterich:2013jsa} 
  C.~Wetterich,
  Variable gravity Universe,
  Phys.\ Rev.\ D {\bf 89}, 
  024005 (2014).
  
\bibitem{Caldwell:1997ii} 
  R.~R.~Caldwell, R.~Dave, and P.~J.~Steinhardt,
  Cosmological imprint of an energy component with general equation of state,
  Phys.\ Rev.\ Lett. {\bf 80}, 
  1582 (1998).
  
\bibitem{trac1} 
  M.~Sami, Models of Dark Energy, in \textit{The Invisible Universe: Dark Matter and Dark Energy}, edited by L. Papantonopoulos, Lecture Notes in Physics Vol. 720 (Springer, Berlin, 
  2007), pp. 219-256.

\bibitem{Ng}
  S.~C.~C.~Ng, N.~J.~Nunes, and F.~Rosati,
  Applications of scalar attractor solutions to cosmology,
  Phys.\ Rev.\ D {\bf 64}, 
  083510 (2001).

\bibitem{Chiba:2009gg} 
  T.~Chiba,
  The equation of state of tracker fields,
  Phys.\ Rev.\ D {\bf 81}, 
  023515 (2010).
 
\bibitem{trac2}
  M.~Sami,
  Dark energy and possible alternatives,
  arXiv:0901.0756 [hep-th].

\bibitem{Lucchin:1984yf} 
  F.~Lucchin and S.~Matarrese,
  Power law inflation,
  Phys.\ Rev.\ D {\bf 32}, 
  1316 (1985).
  
\bibitem{Sami:2002fs} 
  M.~Sami, P.~Chingangbam, and T.~Qureshi,
  Aspects of tachyonic inflation with exponential potential,
  Phys.\ Rev.\ D {\bf 66}, 
  043530 (2002).
  
\bibitem{Ratra:1989uz} 
  B.~Ratra,
  Inflation in an exponential potential scalar field model,
  Phys.\ Rev.\ D {\bf 45}, 
  1913 (1992).

\bibitem{Joyce} 
  M.~Joyce and T.~Prokopec,
  Turning around the sphaleron bound: Electroweak baryogenesis in an alternative post-inflationary cosmology,
  Phys.\ Rev.\ D {\bf 57}, 
  6022 (1998).

\bibitem{Barrow1}
  P.~Parsons and J.~D.~Barrow,
  Generalised scalar field potentials and inflation,
  Phys.\ Rev.\ D {\bf 51}, 
  6757 (1995).

\bibitem{Wetterich}
  J.~Rubio and C.~Wetterich,
  Emergent scale symmetry: Connecting inflation and dark energy,
  Phys.\ Rev.\ D {\bf 96}, 
  063509 (2017).

\bibitem{nonlinear}
  S. Wiggins, \textit{Introduction to Applied Nonlinear Dynamical Systems and Chaos.} (Springer, New York, 1990).
  
\bibitem{Barreiro:1999zs} 
  T.~Barreiro, E.~J.~Copeland, and N.~J.~Nunes, 
  Quintessence arising from exponential potentials,
  Phys.\ Rev.\ D {\bf 61}, 
  127301 (2000). 

\bibitem{Barrow2}
  J.~D.~Barrow and A.~A.~H.~Graham,
  Singular inflation,
  Phys.\ Rev.\ D {\bf 91}, 
  083513 (2015).

\end{thebibliography}
\end{document}